\documentclass[review]{elsarticle}

\usepackage{lineno,hyperref}
\modulolinenumbers[5]

\journal{xxx}

\usepackage{indentfirst}

\usepackage{times}
\usepackage{graphicx} 

\usepackage{subfigure} 
\usepackage{caption}

\biboptions{numbers,sort&compress} 

\usepackage{algorithm}
\usepackage{algorithmic}

\usepackage{amsmath}

\usepackage{multirow}

\usepackage{longtable}

\usepackage{tabularx}

\newcommand{\tabincell}[2]{\begin{tabular}{@{}#1@{}}#2\end{tabular}} %

\usepackage{fancyhdr}

\usepackage{titleref} %

\usepackage{geometry}
\usepackage{url}

\usepackage{threeparttable} 

\usepackage{setspace}

\makeatletter
\newenvironment{breakablealgorithm}
{
	\begin{center}
		\refstepcounter{algorithm}
		\hrule height.8pt depth0pt \kern2pt
		\renewcommand{\caption}[2][\relax]{
			{\raggedright\textbf{\ALG@name~\thealgorithm} ##2\par}%
			\ifx\relax##1\relax 
			\addcontentsline{loa}{algorithm}{\protect\numberline{\thealgorithm}##2}%
			\else 
			\addcontentsline{loa}{algorithm}{\protect\numberline{\thealgorithm}##1}%
			\fi
			\kern2pt\hrule\kern2pt
		}
	}{
		\kern2pt\hrule\relax
	\end{center}
}
\makeatother

\begin{document}

\begin{frontmatter}
	
\title{Transfer-Learning Oriented Class Imbalance Learning for Cross-Project Defect Prediction}	


\author[mymainaddress,mysecondaryaddress]{Haonan Tong}	
\author[mymainaddress,mysecondaryaddress]{Bin Liu}	
\author[mymainaddress,mysecondaryaddress]{Shihai Wang\corref{mycorrespondingauthor}}\cortext[mycorrespondingauthor]{Corresponding author}\ead{wangshihai@buaa.edu.cn}
\author[mymainaddress,mysecondaryaddress]{Qiuying Li}

\address[mymainaddress]{School of Reliability and Systems Engineering, Beihang University, 100191, Beijing, China}
\address[mysecondaryaddress]{Science \& Technology on Reliability \& Environmental Engineering Laboratory, Beihang University, 100191, Beijing, China}

%
%
%
%

\begin{abstract}


Cross-project defect prediction (CPDP) aims to predict defects of projects lacking training data by using prediction models trained on historical defect data from other projects. However, since the distribution differences between datasets from different projects, it is still a challenge to build high-quality CPDP models. Unfortunately, class imbalanced nature of software defect datasets further increases the difficulty. In this paper, we propose a transfer-learning oriented minority over-sampling technique (TOMO) based feature weighting transfer naive Bayes (FWTNB) approach (TOMOFWTNB) for CPDP by considering both class-imbalance and feature importance problems. Differing from traditional over-sampling techniques, TOMO not only can balance the data but reduce the distribution difference. And then FWTNB is used to further increase the similarity of two distributions. Experiments are performed on 11 public defect datasets. The experimental results show that (1) TOMO improves the average G-Measure by 23.7\%$\sim$41.8\%, and the average MCC by 54.2\%$\sim$77.8\%. (2) feature weighting (FW) strategy improves the average G-Measure by 11\%, and the average MCC by 29.2\%. (3) TOMOFWTNB improves the average G-Measure value by at least 27.8\%, and the average MCC value by at least 71.5\%, compared with existing state-of-the-art CPDP approaches. It can be concluded that (1) TOMO is very effective for addressing class-imbalance problem in CPDP scenario; (2) our FW strategy is helpful for CPDP; (3) TOMOFWTNB outperforms previous state-of-the-art CPDP approaches.

\end{abstract}


\end{frontmatter}

\section{Introduction}\label{sec:intro}


Software defect prediction (SDP) has been a significant research topic in the software engineering field for over three decades \cite{Song2011A}. SDP plays an important role in reasonably allocating the testing resources and improving the efficiency of software testing, which has drawn increasing attention of both academic and industrial communities  \cite{Gyimothy2005Empirical,Catal2009A,Kitchenham2009Systematic,Menzies2010Defect,Hall2012A,Shepperd2013Data,Catal20114Software,Ghotra2015Revisting,Rathore2017Linear}. In general, SDP models are built on the historical defect data form a project and then they are used to predict the defects of new software modules (such as classes, methods, and files) from the same project. This kind of SDP is generally called within-project defect prediction (WPDP) \cite{Kanmani2007Object,Elish2008Software,Gao2011Choosing,Dejaeger2013Toward,Czibula2014Software,Yang2015A,Li2016Three,Lee2016Developer,Singn2017Fuzzy,Xu2018Software,Abaei2015An,Rana2015Improving}. However, at the early stage of a project, since there is no sufficient or any historical defect data \cite{Zimmermann2009Cross,Turhan2009On,Peters2013Better}, it is difficult to perform WPDP. 

%

To address this issue, researchers have proposed cross-project defect prediction (CPDP)  \cite{Turhan2009On,Peters2013Better,Ryu2015A,Ma2012Transfer,Xia2016HYDRA,Nam2013Transfer,Jing2015Heterogeneous,Nam2015Heterogeneous,Wang2016Automatically,Ma2017Kernel,Li2018Cost}. CPDP attempts to train a prediction model by using historical defect datasets (aka. source data) collected from other projects and then use the trained model to predict defects of a new project data (aka. target data). Fortunately, there are many defect datasets collected from open source software projects, which are publicly available on PROMISE repository \cite{Menzies2015The}. Some CPDP approaches have been proposed in previous studies. These approaches can be divided into two main categories: instance-based CPDP approaches \cite{Turhan2009On,Peters2013Better,Ryu2015A,Ma2012Transfer,Xia2016HYDRA} and feature-based CPDP approaches \cite{Nam2013Transfer,Jing2015Heterogeneous,Nam2015Heterogeneous,Wang2016Automatically,Ma2017Kernel,Li2018Cost}. Instance-based CPDP approaches demand that source and target datasets have common software metrics. They try to reduce the distribution differences between the source and target datasets by selecting similar instances from source data \cite{Turhan2009On,Peters2013Better,Ryu2015A} or weighting training instances \cite{Ma2012Transfer,Xia2016HYDRA}. Feature-based CPDP (aka. heterogeneous defect prediction) approaches try to find the latent common feature space by feature space transformation \cite{Nam2013Transfer,Jing2015Heterogeneous,Nam2015Heterogeneous,Wang2016Automatically,Li2018Cost}. Especially, CPDP with heterogeneous metric sets between source and target datasets are called as heterogeneous defect prediction \cite{Jing2015Heterogeneous,Nam2015Heterogeneous,Li2018Cost}.




Software defect data has class imbalance nature, where the number of non-defective instances (majority class samples) is much more than that of defective ones (minority class samples) \cite{Boehm1988Understanding}.  As Bohem and Basili \cite{Boehm2001Software} pointed out that in most cases, 20\% of the modules can result in 80\% of the software defects. Class imbalanced distribution is a major reason accounting for poor predicting performance, especially for the minority class \cite{Wang2013Using} \cite{Mahmood2015What}. The class-imbalance problem has been widely studied in WPDP scenario \cite{Aman2015Lines,Chen2015Negative,Aman2015Lines,Laradji2015Software}, and class-imbalance learning approaches used in SDP can be divided into three categories: re-sampling methods \cite{Aman2015Lines}, cost-sensitive learning \cite{Khoshgoftaar2002Cost}, and ensemble learning \cite{Laradji2015Software}. However, most previous CPDP approaches \cite{Turhan2009On,Peters2013Better,Nam2013Transfer,Nam2015Heterogeneous} do not consider the class-imbalance problem. Although a few CPDP methods \cite{Chen2015Negative,Li2018Cost} take it into consideration, they just directly use traditional class imbalance learning approaches (such as synthetic minority over-sampling technique (SMOTE) \cite{Chawla2002SMOTE}, cost-sensitive learning), but do not address the class-imbalance problem in CPDP scenario by considering the distribution characteristics of both source and target datasets.

Different software metrics (aka. features, attributes) usually have different importance when making defect prediction in SDP. Researchers usually measure the importance of each feature according to its relevance with software defect. Feature importance has been widely used in many feature selection methods \cite{Kira1992A,Kannan2010A,Liu2014Two,Liu2014FECAR,Xu2016MICHAC}. Intuitively, the features with high importance should be assigned bigger weights when making a prediction. However, the previous CPDP approaches usually assume that different software metrics have the same importance.

The objective of this paper is to develop an effective cross-project defect prediction modeling approach dealing with the class-imbalance problem. In this paper, we propose a transfer learning oriented over-sampling technique based feature weighting transfer naive Bayes (TOMOFWTNB) approach. To evaluate TOMOFWTNB, we explore the following research questions (RQs):

\begin{itemize}
	\setlength{\itemsep}{-0.08cm}	
	\item RQ1: Does TOMO outperform traditional over-sampling techniques in CPDP scenario?	
	\item RQ2: Is feature weighting strategy helpful for CPDP?
	\item RQ3: Does TOMOFWTNB outperform other CPDP approaches?
	
\end{itemize}

To answer these questions, we perform a series of experiments on 11 public defect datasets from tera-PROMISE repository\footnote{http://openscience.us/repo}. The experimental results show that (1) TOMO is very effective to address the class-imbalance problem in CPDP scenario. (2) our feature weighting strategy is helpful for CPDP. (3) TOMOFWTNB outperforms previous state-of-the-art both instance-based and feature-based CPDP approaches.

The main contributions of this paper are presented as follows:

(1) To address the class imbalance problem in cross-project defect prediction (CPDP) scenario, a transfer learning oriented minority over-sampling approach (TOMO) is proposed by us, which can not only help to balance the data but can increase the similarity of distributions between the source and target data.


(2) To apply the information of feature importance in CPDP scenario, we proposed a feature weighting transfer naive Bayes approach based on previous CPDP approach transfer naive Bayes \cite{Ma2012Transfer}.

(3) Based on the above work, we proposed a novel transfer-learning oriented over-sampling based feature weighting transfer naive Bayes (TOMOFWTNB) approach for CPDP.

(4) We performed experiments on 11 public defect datasets from different software projects. The experimental results show that the proposed TOMO, feature weighting strategy, and our TOFWTNB approach can achieve desirable performance, respectively.

The rest of this paper is organized as follows: Section \ref{sec:RWork} reviews previous work about cross-project defect prediction and class imbalance learning in SDP field. In Section \ref{sec:Methodology}, we present our proposed CPDP approach. The experimental setup and results are presented in Section \ref{sec:experiment_setup} and \ref{sec:experiment_results}, respectively. Our research is discussed in Section \ref{sec:discussion}. Some potential threats to our study are shown in Section \ref{sec:threats}. Finally, we make a conclusion on our research in Section \ref{sec:conclusion}.

\section{Related Works}\label{sec:RWork}
In this section, we briefly introduce existing cross-project defect prediction models and class imbalance learning methods in software defect prediction field.

\subsection{Transfer Learning}
A major assumption in many machine learning approaches is that the training and future data must have same feature space and the same distribution \cite{Pan2010A}. Transfer learning techniques do not require this assumption and allow transfer domain knowledge between different domains. According to \cite{Pan2010A}, the definition of transfer learning is: \emph{Given a source domain $D_S$ and learning task $T_S$, a target domain $D_T$ and learning task $T_T$, transfer learning aims to help improve the learning of the target predictive function $f_T\left(\cdot\right)$ in $D_T$ using the knowledge in $D_S$ and $D_S$, where $D_S\neq D_T$, or $T_D\neq T_T$.}

Transfer learning techniques have been applied in many fields, such as image classification \cite{Quattoni2008Transfer}, speech recognition \cite{Wang2015Transfer}, face recognition \cite{Cao2013A}, robots modeling \cite{Um2014Independent}, and recommended systems \cite{Grolman2016Utilizing}. Actually, cross-project defect prediction is also a transfer learning problem in the SDP field \cite{Ma2012Transfer}.

\subsection{Cross-project Defect Prediction}
Recently, CPDP has drawn increasing attention of academic and industrial communities, and many CPDP approaches have been proposed.

Zimmermann et al. conducted 622 cross-project predictions and the results suggested that training data selection plays an important role in successful CPDP \cite{Zimmermann2009Cross}. Turhan et al. proposed nearest-neighbor (NN) based filter to select training instances from source data which are regarded as nearest-neighbors of target data \cite{Turhan2009On}. Peters et al. presented Peters filter to select training data via the structure of data from other similar projects \cite{Peters2013Better}. Ryu et al. developed a nearest-neighbor based hybrid instance selection (HISNN) method for CPDP \cite{Ryu2015A}. Differing from above approaches which select source instances similar to target data to train classifiers, some models attempt to reduce the distribution difference between the source and target data by weighting source instances \cite{Ma2012Transfer,Xia2016HYDRA}. Ma et al. proposed a CPDP approach named transfer naive Bayes (TNB), which extended naive Bayes with weighted instances \cite{Ma2012Transfer}. For each instance in source data, the similarity weights are calculated based on the distributional characteristics of target data and then constructed a classifier based on naive Bayes. Xia et al. developed a CPDP approach named HYDRA, which adjusted the weights of source instances during the training of the extended AdaBoost algorithm \cite{Xia2016HYDRA}. 

Above CPDP approaches suppose that source and target datasets have common software metrics (aka. features or attributes). These approaches try to reduce the distribution differences between the source and target datasets by selecting similar instances or weighting instances. This kind of approaches can be called instance-based CPDP methods. For the case that source and target datasets have no common features or have a different number of metrics, some CPDP approaches called feature-based CPDP or heterogeneous defect prediction approaches have been proposed \cite{Nam2013Transfer} CCA+ \cite{Jing2015Heterogeneous} HDB-KS \cite{Nam2015Heterogeneous} DBN \cite{Wang2016Automatically} KCCA+ \cite{Ma2017Kernel} CSKCCA+ \cite{Li2018Cost}. Nam et al. used extended transfer component analysis by using normalization techniques (TCA+) to find latent common feature space for CPDP \cite{Nam2013Transfer}. Nam and Kim applied metric selection and metric matching to develop a CPDP model. Jing et al. firstly applied canonical correlation analysis (CCA+) for CPDP \cite{Jing2015Heterogeneous}. To improve the performance of CCA+ in the nonlinear scenario, Ma et al. proposed a CPDP approach named kernel canonical correlation analysis (KCCA+) \cite{Ma2017Kernel}. Considering the class imbalance problem in CPDP, Li et al. combined cost-sensitive learning and KCCA+ and proposed cost-sensitive transfer kernel canonical correlation analysis (CTKCCA) approach \cite{Li2018Cost}.

%
%
%
%
%
%
%
%
%
%
%
%
%
%
%


Previous works have proposed many methods for CPDP. However, the class imbalance problem is rarely taken into consideration, which may limit the prediction performance. Furthermore, previous CPDP approaches usually assume that different metrics have the same importance for predicting software defect proneness. Actually, different metrics usually have distinct importance since they usually have a different contribution for predicting software defect. The most intuitive evidence is that the correlations between different metrics and software defects are usually different. Therefore, software metrics should be treated differently according to their importance.

\subsection{Class Imbalance Learning for Software Defect Prediction}


To address the class-imbalance problem in SDP, many class-imbalance learning methods have been used at data and algorithm levels \cite{Li2018Cost,Laradji2015Software}. 

Data-level methods refer to a variety of re-sampling techniques, which try to balance datasets by removing majority class samples or adding minority class ones. Many re-sampling methods have been used for SDP, such as random under-sampling, random over-sampling, SMOTE \cite{Chawla2002SMOTE}, and their variants \cite{Aman2015Lines,Chen2015Negative,Siers2015Software,Gao2014The,Khoshgoftaar2014A,Zhang2016Label}. However, this kind of methods may have the risk of discarding useful information or introducing data noise.


Algorithm-level methods mainly include ensemble learning approaches and cost-sensitive learning. Classic ensemble learning techniques, such as bagging and boosting, have demonstrated the effectiveness in handling class imbalanced problem although not specifically designed to address this problem \cite{Galar2012A}. Many variants of classic ensemble learning approaches have been proposed to address the class imbalance problem in the SDP field \cite{Sun2012Using,Wang2013Using,Laradji2015Software,Tong2018Software}. However, how to effectively design an appropriate integration mechanism is still not be completely solved. Cost-sensitive (CS) learning addresses class imbalance problem by assigning big misclassification cost for defective instance and small misclassification cost for non-defective instances. Khoshgotaar et al. \cite{Khoshgoftaar2002Cost} first introduced cost-sensitive learning into SDP by proposing a cost-boosting method. Zheng \cite{Zheng2010Cost} proposed cost-sensitive boosting neural networks for SDP. Similarly, CS neural network was studied by Arar and Ayan \cite{Arar2015Software}. Liu et al. \cite{Liu2014Two} proposed a two-stage CS learning for SDP, which includes CS feature selection and CS neural network classifier. Li et al. \cite{Li2016Three} used three-way decision-based CS for SDP. Some researchers combined CS with other machine learning methods, such as dictionary learning and random forest \cite{Jing2014Dictionary,Siers2015Software}. Furthermore, CS also have been used in the CPDP scenario \cite{Ryu2017A,Li2018Cost}. However, how to set suitable cost values is still an unsolved problem for the cost-sensitive learning method.



Although these class-imbalance learning methods have been demonstrated to be effective in WPDP scenario and even some of them has been used recently in CPDP scenario \cite{Ryu2017A,Li2018Cost}, these methods are not specially designed for addressing the class-imbalance problem in CPDP scenario. More specifically, these methods do not concern the distribution differences between the source and target datasets in CPDP scenario. Take SMOTE as an example, the artificial minority class samples are generated based on the real minority class samples in source dataset. Thus, the artificial minority class samples cannot ensure that the expended source dataset by adding the artificial samples even increase the distribution difference between the source and the target datasets. Differing from above class imbalance learning methods, we propose a transfer-learning oriented minority over-sampling technique (TOMO, see Section \ref{sec:TOMO}) to generate synthetic minority class samples by considering both source and target data distribution information. In this way, our synthetic minority class samples not only balance the dataset but also increase the similarity between source and target datasets.

\section{Proposed Method}\label{sec:Methodology}
In this section, we describe the proposed transfer-learning oriented minority over-sampling technique based feature weighting transfer naive Bayes (TOMOFWTNB) approach.

\begin{figure}[htbp]
	\begin{center}
		\centering\includegraphics[width=\columnwidth]{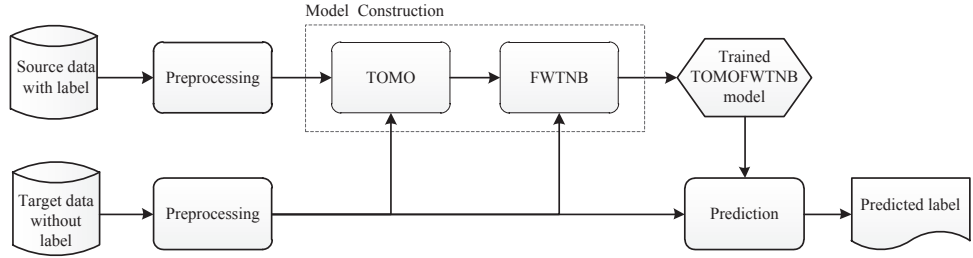}
		\caption{Overview framework of proposed TOMOFWTNB for cross-project defect prediction}
		\label {fig:framework}
	\end{center}
\end{figure}

Fig \ref{fig:framework} presents the overall framework of proposed TOMOFWTNB for CPDP. TOMOFWTNB assumes that (1) source dataset and target dataset has common software metrics; (2) source data is imbalanced. Given a labeled source dataset and a target dataset without the labels, we first preprocess the data. For a source instance, if the number of defects is larger than zero, then we label it as 1, otherwise 0. And then, We remove the instances having missing values and the duplicate instances. Secondly, we balance the class-imbalanced source data by using our proposed transfer learning oriented minority over-sampling technique (TOMO). And then we build a CPDP model by using our proposed FWTNB approach. Finally, we use the trained model (TOMOFWTNB) to predict the label of target data and evaluate its performance.

The proposed TOMOFWTNB approach includes two key phases: TOMO and FWTNB. In the following paragraphs, we present the details of these two phases, respectively.

\subsection{Transfer-learning Oriented Minority Over-sampling Technique} \label{sec:TOMO}

Since traditional class-imbalance learning approaches (e.g., SMOTE) widely used in WPDP scenario do not consider the distribution difference between the source and target datasets, they are not very suitable for CPDP scenario. To address class imbalance problem in CPDP scenario, we propose a novel transfer learning oriented minority over-sampling technique (TOMO). TOMO takes both the distribution characteristics of source data and that of target data into consideration. Thus, TOMO can not only balance the source data but reduce the distribution difference between the source and target data. Algorithm \ref{alg:TOMO} shows the Matlab pseudo-code of TOMO.   

Given source data $S$ and target data $T$, the minority and majority class sample subsets are denoted by $S_P$ and $S_N$, respectively. TOMO includes two main stages: (1) construct the nearest-neighbor matrix; and (2) generate artificial minority samples.

\begin{figure}[htbp]
	\begin{center}
		\centering\includegraphics[width=\columnwidth]{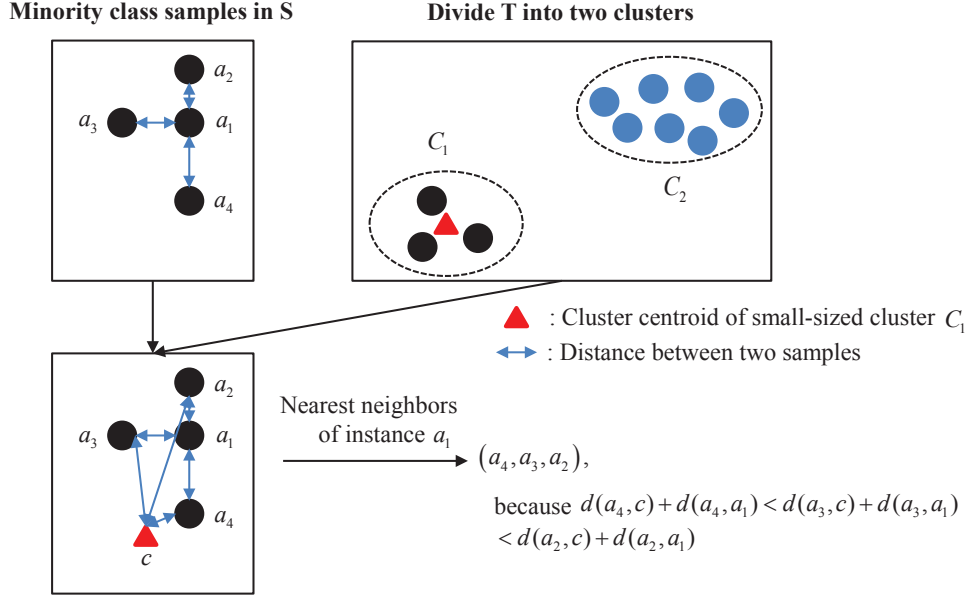}
		\caption{Notation illustration for finding nearest-neighbors in Algorithm \ref{alg:TOMO}. If just consider S, then the nearest neighbors of instance $a_1$ is $\left\{a_2, a_3, a_4\right\}$. But if consider both S and T, the nearest neighbors of instance $a_1$ is $\left\{a_4, a_3, a_2\right\}$ where $\lambda$=0.5.
		}
		\label {fig:nearNei}
	\end{center}
\end{figure}

In the first stage, the target data $T$ is divided into two clusters in which the small-sized cluster is regarded as the \emph{potential} minority class samples and its centroid is denoted as $c$. And then, all instances in $S_P$ are sorted in ascending order according to their distance from cluster centroid $c$. For each instance in sorted $S_P$, the order of its nearest-neighbors (i.e., other instances in $S_P$) are next determined by both the distance from its every neighbor to current instance and the distance from its every neighbor to the cluster centroid $c$ of potential minority class samples in $T$. In order to be able to flexibly adjust the importance of above two distances, a parameter $\lambda$ (Step 11) is introduced. The process of finding nearest-neighbor for an instance can be briefly illustrated as Fig \ref{fig:nearNei}. 

In the second stage, artificial minority class samples are generated based on the above nearest-neighbor matrix. For convenience, let the number of artificial minority class samples be $n_0$, and the size of $S_P$ be $n_P$. According to the size of $n_0$, our sampling procedure includes three specific cases. The first case, if $n_0<n_P$, we just use the first $n_0$ samples in $S_P$ and their corresponding first nearest-neighbor to generate artificial minority class samples (Lines 16-19). The second case, if $n_P\leq n_0\leq n_p\left(n_P-1\right)$, we first use each sample in $S_P$ and its corresponding the first $floor(n_0/n_P)$ nearest-neighbors to generate one part of artificial samples, and then the rest is generated based the first several samples in $S_P$ and their corresponding the first nearest-neighbor (Lines 21-34). Otherwise, we first generate one part of artificial minority class samples by using each sample in $S_P$ and its every nearest-neighbor multiple times and then perform a similar procedure like second case (Lines 36-59).

\begin{breakablealgorithm}
	\caption{$TOMO(S,T,Ratio,\lambda)$}
	\label{alg:TOMO}
	\begin{algorithmic}[1] 
		\STATE {\bfseries Input:} 
		\STATE (1) $S$: Source dataset
		\STATE (2) $T$: Target dataset
		\STATE (3) $Ratio$: Ideal imbalance ratio $Ratio$ i.e., (Size of original and synthetic minority class)/(Size of majority class)
		\STATE (4) $\lambda$: Weight coefficient with value range [0,1], by default $\lambda=0.4$ 
		
		\STATE {\bfseries Program Begin}
		
		\STATE Divide $T$ into two clusters and regard samples in the smaller cluster with centroid $C_{min}$ as potential minority class samples; denote the majority and minority class samples in $S$ by $S_N$ and $S_P$, respectively. 	
		
		\STATE Sort $S_P$ in ascending order according to the distance (Euclidean distance, by default) between $C_{min}$ and every instance in $S_P$ and then construct a distance matrix $distSS$ in which $distSS(i,j)$ denotes the distance between $i$-th and $j$-th samples in sorted $S_P$;
		
		\STATE Calculate the distance between $C_{min}$ and each sample in sorted $S_P$ and can obtain a row vector, and then replicate it $n_P-by-1$ times to generate a matrix $distTS$;  
		
		\STATE Normalize $distSS$ and $distTS$ as: $distSS(i,j)=\frac{distSS(i,j)}{sum(distSS(i,:))}$, $distTS(i,j)=\frac{distTS(i,j)}{sum(distTS(i,:))}$;
		
		\STATE Calculate hybrid distance matrix $distH$:
		$distH=\lambda\ast distSS_N+\left(1-\lambda\right)distTS_N$;
		
		\STATE Construct the index matrix $NeigInd$ of nearest-neighbors for each sample of sorted $S_P$ by sorting $distH$ in ascending order by row.
		$NeigInd(i,j)$ denotes the index of $j$-th nearest neighbor of $i$-th minority class sample. 
		
		
		\STATE The number of synthetic minority class samples $synthetic$ :$n_0=floor(n_N\ast Ratio)-n_P$, where $n_N$ and $n_P$ denote
		\STATE $k=n_0/n_P$.
		
		\IF{$k<1$}
			\FOR{$i=1:n_0$}
			\STATE Generate a random number $rand\in\left(0,1\right)$. 
			\STATE 
			$synthetic(i,:)=S_P(i,:)-rand\ast(S_P(NeigInd(i,1),:)-S_P(i,:))$
			\ENDFOR
		
		\ELSIF{$k>=1$ and $k<=(n_P-1)$} 
			\STATE $j_0=1$.
			\FOR{$i=1:n_P$}
				\FOR{$j=1:floor(k)$}
					\STATE $j_0=j_0+1$; 
					\STATE Generate a random number $rand\in\left(0,1\right)$;
					\STATE $synthetic(j_0,:)=S_P(i,:)-rand\ast(S_P(NeigInd(i,j),:)-S_P(i,:))$;
				\ENDFOR
			\ENDFOR
			
			\STATE Let $ind=1:(n_0-floor(k)\ast n_P)$;
			\FOR{$i=1:length(ind)$}
				\STATE $j_0=j_0+1$; 
				\STATE Generate a random number $rand\in\left(0,1\right)$; 
				\STATE  $synthetic(j_0,:)=S_P(ind(i),:)-rand\ast(S_P(NeigInd(ind(i),floor(k)+1),:)-S_P(ind(i),:))$;
			\ENDFOR
		
		\ELSE
			\STATE $k_0=floor(k)/(n_P-1)$;
			\STATE $j_0=0$.
			\FOR{$i_0=1:floor(k_0)$}
				\FOR{$i=1:1:n_P$}
					\FOR{$j=1:n_P-1$}
						\STATE $j_0=j_0+1$; 
						\STATE Generate a random number $rand\in\left(0,1\right)$; 
						\STATE $synthetic(j_0,:)=S_P(i,:)-rand\ast(S_P(NeigInd(i,j),:)-S_P(i,:))$
					\ENDFOR
				\ENDFOR
			\ENDFOR
			
			\FOR{$i=1:n_P$}
				\FOR{$j=1:(floor(k)-floor(k_0)\ast(n_P-1))$}
					\STATE $j0=j0+1$; 
					\STATE Generate a random number $rand\in\left(0,1\right)$; 
					\STATE $synthetic(j0,:)=S_P(i,:)-rand\ast(S_P(NeigInd(i,j),:)-S_P(i,:))$;
				\ENDFOR
			\ENDFOR
			
			\STATE Let $ind=1:(n_0-floor(k)\ast n_P)$;
			\FOR{$i=1:length(ind)$}
				\STATE $j_0=j_0+1$; 
				\STATE Generate a random number $rand\in\left(0,1\right)$; 
				\STATE $synthetic(j_0,:)=S_P(ind(i),:)-rand\ast(S_P(NeigInd(ind(i),floor(k)-(floor(k_0))\ast (n_P-1)+1),:)-S_P(ind(i),:))$;
			\ENDFOR
		
		\ENDIF
		\STATE{\bfseries End of Program}
		
		\STATE {\bfseries Output:} Synthetic minority class samples $synthetic$.	
	\end{algorithmic}
\end{breakablealgorithm}  

\subsection{Feature Weighting Transfer Naive Bayes: FWTNB}

\subsubsection{Feature importance based on Maximal Information Coefficient}

Feature importance can be identified according to the relevance between feature and software defect according to previous SDP studies \cite{Liu2014Two,Xu2016MICHAC,Liu2014FECAR}. Many indicators have been used to measure the relevance, such as Pearson's correlation coefficient and mutual information. However, Pearson's correlation coefficient is not suitable for measuring non-linear relationship, mutual information is sensitive to discretization methods when discretizing continue variables before calculating mutual information. Actually, the relationship between software metrics and the class label is not always the simple linear relationship, and many software metrics are continuous variables. To solve this problem, we used maximal information coefficient (MIC) to measure the relevance between software metrics and the class label (i.e., defective or not).


MIC is a robust indicator of measuring the correlation of two variables, which was developed by Reshef et al in 2011 \cite{David2011Detecting}. Xu et al. \cite{Xu2016MICHAC} proposed a feature selection method named MICHAC for software defect prediction, in which MIC was used to measure the correlation between software metrics and the class label. 

%
%


In this study, we use a MATLAB API named \textbf{minepy} developed by Albanese et al. \cite{Albanese2013Minerva} to calculate MIC and apply this API with its default parameters. 

\subsubsection{Weighting for source data based on data similarity}

%
%

To reduce the distribution difference between the source and target datasets, Ma e al. \cite{Ma2012Transfer} proposed a method based on sample similarity and data gravitation to allocate a weight for each instance in source dataset. However, the similarity between each source instance and target dataset is calculated by assuming that software metrics have the same contribution to predict the class label. Therefore, in this paper, we propose feature weighting similarity.

Suppose an instance without the class label can be represented as $x_i=\left\{a_{i1},a_{i2},\cdots,a_{ik}\right\}$, where $k$ denotes the number of software metrics. The maximum and minimum values for $j$-th metric in target dataset are:

\begin{equation}
max_j=max\left\{a_{1j},a_{2j},\cdots,a_{nj}\right\}
\end{equation}
\begin{equation}
min_j=min\left\{a_{1j},a_{2j},\cdots,a_{nj}\right\}
\end{equation}

And then for each instance $x_i$ in training dataset, the \textbf{feature weighting similarity} of $x_i$ to the target data is calculated as:

\begin{equation}
s_i=\sum_{i=1}^kh\left(a_{ij}\right)\frac{MIC_i}{\sum_{i=1}^kMIC_i}
\end{equation}
where $h\left(a_{ij}\right)=\left\{\begin{array}{lc}1,&if\;min_j\leq a_{ij}\leq max_j\\0,&otherwise\end{array}\right.$, and $MIC_i$ denotes the MIC value of $i$-th metrics.

According to the formulation of data gravitation \cite{Peng2009Data}, weight $w_i$ of $i$-th instance in training dataset can be defined as:

\begin{equation}
w_i=\frac{s_i}{(\sum_{i=1}^kMIC_i-s_i+1)^2}
\end{equation}

\subsubsection{Feature weighting Prediction}

Firstly, the prior probability of class is calculated based on weighted source data. According to \cite{Frank2002Locally}, the weighted prior probability for given class $c$ can be calculated as follows:

\begin{equation}
P(c)=\frac{\sum_{i=1}^nw_i\sigma\left(c_i,c\right)+1}{\sum_{i=1}^nw_i+n_c}
\end{equation}
where $c_i$ denotes the actual class of $i$-th source instance, $n$ represents the number of instances in source data, $n_c$ means the number of different classes ($n_c=2$ in this study), $\sigma(x,y)$ is an indicator function, i.e., $\sigma(x,y)=1$ if $x=y$, otherwise $\sigma(x,y)=0$.

Given a target instance $x_m$, according to \cite{Frank2002Locally}, the class-conditional probability of $j$-th feature $a_j$ in $x_m$ is calculated as follows
 
\begin{equation}
P(a_j\left|c\right.)=\frac{\sum_{i=1}^nw_i\sigma\left(a_{ij},a_j\right)\sigma\left(c_i,c\right)+1}{\sum_{i=1}^nw_i\sigma\left(c_i,c\right)+n_j}
\end{equation}
where $a_{ij}$ represents the value of $j$-th feature of $i$-th source instance, $n_j$ denotes the number of different values of $j$-th feature in source data.

If different features are thought to have same importance, as Ma et al. done in \cite{Ma2012Transfer}, the label of a target instance $u=\left\{a_1,a_2,\cdots,a_k\right\}$ can be predicted as follows:

\begin{equation}
h(u)=\underset{c\in C}{arg\;maxP(c\left|u\right.)}=arg\;\underset{c\in C}{max}\frac{P(c)\prod_{j=1}^kP(a_j\left|c\right.)}{\sum_{c\in C}P(c)\prod_{j=1}^kP(a_j\left|c\right.)}
\end{equation}
where $a_j$ denotes the value of $j$-th feature of the target instance $u$, $k$ represents the number of features, $P(c)$, $P(a_j\left|c\right.)$ and $P(c\left|u\right.)$ denotes the prior, class-conditional, and posterior probability of class $c$.

However, since the correlation degree between different features and the class label is usually different, they should be treated differently. To this end, we propose the \textbf{feature weighting posterior probability} in the CPDP scenario as follows:

\begin{equation} \label{eq:fw_postProb}
h(u)=\underset{c\in C}{arg\;maxP(c\left|u\right.)}=arg\;\underset{c\in C}{max}\frac{P(c)\prod_{j=1}^k\left(P(a_j\left|c\right.)\right)^{exp\left(\frac{MIC_j}{\sigma^2\sum_{i=1}^kMIC_i}\right)}}{\sum_{c\in C}P(c)\prod_{j=1}^k\left(P(a_j\left|c\right.)\right)^{exp\left(\frac{MIC_j}{\sigma^2\sum_{i=1}^kMIC_i}\right)}}
\end{equation}
where $MIC_j$ denotes the MIC value between $j$-th feature and the class label in source data, $\sigma$ is positive hyper-parameter, by default $\sigma=1$ in this study, $P(c)$, $P(a_j\left|c\right.)$ and $P(c\left|u\right.)$ denotes the prior, class-conditional, and posterior probability of class $c$, respectively.

To eliminate the effect of data magnitude, we preprocess original values of all features in source and target data by using logarithm transformation. Since the features of software defect data are usually numeric, just like Ma et al. did in \cite{Ma2012Transfer}, we use Fayyad and Irani’s minimum description length (MDL) based discretization method \cite{Fayyad1993Multi} to discretize all numeric features after calculating feature importance.

\section{Experimental Setup}\label{sec:experiment_setup}

%

\subsection{Research Questions} \label{subsec:RQ}

\begin{itemize}
	\item \textbf{RQ1: Does TOMO outperform traditional over-sampling techniques in CPDP scenario?}
	
	\item \textbf{RQ2: Is feature weighting strategy be helpful for CPDP?}
	
	\item \textbf{RQ3: Does TOMOFWTNB outperform other CPDP approaches?}
	
	
\end{itemize}

To answer RQ1, we compare TOMO with one of the most commonly used traditional over-sampling technique, i.e., synthetic minority over-sampling technique (SMOTE) \cite{Chawla2002SMOTE} in SDP. SMOTE has a most important parameter $N$ which denotes the minority class is over-sampled at N\% of its original size \cite{Chawla2002SMOTE}. Therefore, we implement SMOTE($N$) with different parameters, i.e., SMOTE(100), SMOTE(200), SMOTE(300), SMOTE(400), and SMOTE(500). After over-sampling, we then use transfer naive Bayes (TNB) as common CPDP classifier. TNB is an instance-based CPDP approach proposed by Ma et al\cite{Ma2012Transfer}.

To answer RQ2, we compare the proposed FWTNB with TNB \cite{Ma2012Transfer}. Actually, FWTNB is a feature weighting version of TNB. Since TNB is subject to the imbalanced defect dataset, SMOTE(100) is used to preprocessed the source data before performing FWTNB and TNB.

To answer RQ3, we compare our proposed TOMOFWTNB with \textbf{seven} existing CPDP approaches including Burak Filter \cite{Turhan2009On}, TNB \cite{Ma2012Transfer}, TCA+ \cite{Nam2013Transfer}, CCA+ \cite{Jing2015Heterogeneous}, HISNN \cite{Ryu2015A}, DBN-CP \cite{Wang2016Automatically}, and CTKCCA \cite{Li2018Cost}. The baselines include instance-based CPDP approaches (such as Burak Filter, TNB) and feature-based CPDP approaches (such as CCA+ and CTKCCA).

\subsection{Datasets}

\textbf{Eleven} software defect datasets from tera-PROMISE repository\footnote{http://openscience.us/repo} created by Menzies et al. \cite{Menzies2015The} were used in this study. These datasets were collected by Jureczko and Spinellis \cite{Jureczko2010Using} from different object-oriented software projects. Each instance corresponds to a software class (module) and includes two parts: values of 20 class-level software metrics (such as CK metrics \cite{Chidamber1994A} and QMOOD metrics \cite{Bansiya2002A}) and the number of defects. If a module has one or more defects, then we think this module is defective. Otherwise, we think this module is non-defective. The details of 20 metrics can be seen in \cite{Jureczko2010Using}. 


These datasets were selected based on the following \textbf{three criteria}:(1) The dataset must be \emph{publicly available} so that our research can be more easily verified. (2) All datasets must have \emph{common software metrics} since it is one of our model assumptions (see Section \ref{sec:Methodology}). (3) For the multi-version datasets, only \emph{one version} was used since the distribution of two versions for a project may be very similar even identity.  

Table \ref{tb:statistics} presents the statistics of each of the eleven datasets used in this study. For each dataset, statistical information includes name, metric level, number of software metrics, number of instances, number of defective instances, and defective rate (i.e., the number of defective instances is divided by the total number of instances).

\begin{table}[htbp]
	\centering
	\setlength{\abovecaptionskip}{0pt}
	\setlength{\belowcaptionskip}{10pt}
	\renewcommand{\arraystretch}{1.0} 
	\small
	\caption{\sc{Statistics of the 11 Datasets Used in this Study}}
	\label{tb:statistics}
	\begin{tabular}{lccccc}
		\hline
		\multirow{2}{*} {\textbf{Dataset}} & \multirow{2}{*} {\textbf{Metric Level}} & \multirow{2}{*} {\textbf{\# Metrics}} & \multirow{2}{*} {\textbf{\# Instances}} & \multirow{2}{*} {\textbf{\tabincell{c}{\# Defective\\ Instances}}} & \multirow{2}{*} {\textbf{\tabincell{c}{ Defective\\ Rate}}} \\ \\
		\hline
		ant-1.7 & class & 20 & 745 & 166 & 0.2228 \\
		camel-1.0 & class & 20 & 339 & 13 & 0.0383 \\
		ivy-1.4 & class & 20 & 241 & 16 & 0.0664 \\
		jedit-4.0 & class & 20 & 306 & 75 & 0.2451 \\
		log4j-1.0 & class & 20 & 135 & 34 & 0.2519 \\
		poi-2.0 & class & 20 & 314 & 37 & 0.1178 \\
		prop-6 & class & 20 & 660 & 66 & 0.1000 \\
		tomcat & class & 20 & 858 & 77 & 0.0897 \\
		velocity-1.6 & class & 20 & 229 & 78 & 0.3406 \\
		xalan-2.4 & class & 20 & 723 & 110 & 0.1521 \\
		xerces-1.2 & class & 20 & 440 & 71 & 0.1614 \\
	  
		\hline
	\end{tabular}
\end{table}

\subsection{Performance Evaluation Measures}\label{sunset:perfMeasure}

To evaluate the performance of our proposed model built on imbalanced software defect datasets, two widely used overall performance in previous studies \cite{Peters2013Better,Chen2015Negative,He2013Learning,Peters2013Balancing,Zhang2016Towards,Bowes2016Mutation}, i.e., G-measure and Matthews correlation coefficient (MCC).

For convenience, we first calculate the confusion matrix \cite{Shepperd2014Researcher} as shown in Table \ref{tb:confmat}:

\begin{table}[htbp] %
	\setlength{\abovecaptionskip}{-5pt}
	\setlength{\belowcaptionskip}{0pt}
	\renewcommand{\arraystretch}{1.5}
	\caption{Confusion Matrix} 
	\label{tb:confmat}
	\begin{center}
		\begin{small}
			\begin{tabular}{l|l|l|l}
				\hline
				\multicolumn{2}{c|}{} & \multicolumn{2}{c}{\textbf{Predicted}} \\ 
				\cline{3-4}
				\multicolumn{2}{c|}{} & Positive & Negative \\ 
				\hline
				
				\multirow{2}{*}{\textbf{Actual}} & \multicolumn{1}{c|}{Positive}  & \emph{TP} & \emph{FN} \\ 
				\cline{2-4}
				\multicolumn{1}{l|}{} & Negative & \emph{FP} & \emph{TN} \\ 
				\hline
			\end{tabular}
		\end{small}
	\end{center}
\end{table}


By convention, the defective modules are regarded as positive class samples and the non-defective modules as negative class ones \cite{Wang2013Using}. And then, our performance measures can be defined as follows:

\textbf{The probability of detection (PD)}, aka. recall, true positive rate (TPR), or sensitivity, refers to the proportion of the number of correctly classified positive instances to the total number of actually positive instances:

\begin{equation}
PD=recall=TPR=\frac{TP}{TP+FN}
\end{equation}
 
\textbf{The probability of false alarm (PF)}, aka. false positive rate (FPR), is the proportion of the number of actual negative instances which are misclassified as positive instances to the total number of actually negative instances:
 
\begin{equation}
PF=FPR=\frac{FP}{FP+TN}
\end{equation}


\textbf{G-Measure} is the harmonic mean of PD and (1-PF), which has been widely used in previous SDP studies \cite{Peters2013Better,Chen2015Negative,He2013Learning,Peters2013Balancing,Zhang2016Towards}. G-Measure is a good indicator of performance for imbalanced. G-Measure is defined as:
\begin{equation}
G-Measure=\frac{\left(2\ast PD\ast\left(1-PF\right)\right)}{PD+(1-PF)}
\end{equation} 


\textbf{Matthews correlation coefficient (MCC), Matthews1975Comparison} represents the correlation coefficient between the actual and predicted values in binary classification. It is an overall performance measure for taking $TP$, $TN$, $FP$, and $FN$ into consideration. MCC has been widely used in previous SDP studies \cite{Chen2015Negative,Jing2015Heterogeneous,Zhang2016Towards,Bowes2016Mutation} since it can be utilized even the data is unbalanced \cite{Shepperd2014Researcher}. The range of MCC value is [-1,1], where 1 denotes a perfect prediction, and -1 indicates complete disagreement between actual and predicted values. MCC also has been widely used in previous SDP studies. MCC is defined as:

\begin{equation}
\label{eq:mcc}
MCC=\frac{TP\ast TN-FP\ast FN}{\sqrt{(TP+FP)(TP+FN)(TN+FP)(TN+FN)}}
\end{equation}

Other measures such as accuracy, precision, and F-Measure are not used in this study since they are poor performance measures for data where the defective instances are rare according to \cite{Peters2013Balancing}. Moreover, according to \cite{He2009Learning,Davis2006The}, the area under the ROC curve (AUC) may fail to properly capture the classifier behavior for overly imbalanced data. Therefore, AUC is also not used in this paper.

With respect to PD, PF, or G-Measure, the range of values is [0,1]. A good prediction model should have big PD, G-Measure, and MCC, but small PF. Note that, when all positive instances are misclassified ($TP$=0) and all negative instances are correctly classified ($FP$=0), the value of MCC is NAN because of $TP$+$FP$=0. To avoid this situation, we just replace $FP$=0 with $FP$=1. And then, the value of MCC is zero.

\subsection{Statistic Test}\label{subset:sta_test}

To compare our proposed TOMOFWTNB with existing CPDP approaches, we perform \textbf{Wilcoxon rank-sum test} \cite{Wilcoxon1945Individual} and measure effect size by using \textbf{Cliff's delta} \cite{Cliff1996Ordinal}. They have been commonly used in previous SDP studies \cite{Lee2016Developer,Yang2015Deep,Zhang2016Towards}.

Wilcoxon rank-Sum test is an effective non-parametric statistical test method to check whether the differences between two independent sample sets are statistically significant (which happens when the p-value is smaller than the significance level). \textbf{Wilcoxon rank-Sum test at 5\% significance level} is used in this paper.


Cliff’s delta is a non-parametric effect size measure to quantify the difference between two group samples. Cliff’s delta can be calculated as the method proposed by Macbeth et al \cite{Macbeth2011Cliff}. The value of Cliff's delta ranges from -1 to 1, where -1 or 1 means all values in the first group are smaller or bigger than the second group, and 0 indicates the values in two group are similar. Table \ref{tb:cliff_delta} shows the mappings of Cliff’s delta and effect size.

\begin{table}[htp] %
	\setlength{\abovecaptionskip}{-5pt}
	\setlength{\belowcaptionskip}{0pt}
	\caption{ Mappings of Cliff's Delta to Effect size \cite{Cliff1996Ordinal}}
	\label{tb:cliff_delta} 
	\begin{center}
		\begin{small}
			\begin{tabular}{cc}
				\hline
				\textbf{Cliff's Delta $(\delta)$}  & \textbf{Effect Size Level} \\
				\hline
				$\left|\delta\right|<0.1470$ & Negligible \\
				$0.1470\leq\left|\delta\right|<0.3300$ & Small \\
				$0.3300\leq\left|\delta\right|<0.4740$ & Medium \\
				$\left|\delta\right|\geq0.4740$ & Large \\ 
				\hline
			\end{tabular}
		\end{small}
	\end{center}
\end{table}


\subsection{Experimental Settings}

\subsubsection{Evaluation Settings}
We utilized 11 public defect datasets having a different degree of defective rate (DR) in this study. According to DR, the first seven datasets with the smallest DR were respectively selected as source data, and the first five datasets with the largest DR were respectively select as target data. And then, total \textbf{34} (7*5-1) two-tuples (i.e., Source$\Rightarrow$Target) were obtained after removing one two-tuple (xerces-1.2$\Rightarrow$xerces-1.2). The left side of $\Rightarrow$ indicates the source project dataset, and the right side of $\Rightarrow$ represents the target project dataset.

Given source dataset and target dataset (i.e., one of 34 two-tuples), we repeat each CPDP approach \textbf{30 times}. For each time, as \cite{Li2018Cost}, we randomly select \textbf{90\%} instances of source dataset as the actual training data (i.e., test data) and take all instances of target dataset as test data. Finally, we report the mean performance for each target dataset.

\subsubsection{Parameter Settings}

%

For TOMO, FWTNB, and TOMOFWTNB, we just use our default values for all hyper-parameters, i.e., $Ratio=1$, $\lambda=0.4$, $\sigma=1$. Since we expect the number of minority class samples in the source dataset is equal to that of majority class samples after balancing source dataset with TOMO, we set $Ratio$ to 1. For $\lambda$ and $\sigma$, the rationality of their default values used in this paper and their effect on the performance of TOMOFWTNB are discussed in Section \ref{sec:discussion}. 


For baseline CPDP approaches, we set the number of nearest-neighbors to 5 as \cite{Jing2015Heterogeneous} for Burak filter. For CCA+, since the authors do not state that how to determine the best dimension of projected space or specify the dimension value used in their study, we just set the dimension to 10. For DBN-CP, we set the range of the number of layers to $\left\{3,5\right\}$, the range of the number of nodes to $\left\{20,50,100\right\}$, and the number of iteration to 50. For other approaches, we use the same parameters as the authors adopted in their studies. Note that, with respect to CTKCCA, we just use the source code\footnote{ https://sites.google.com/site/cstkcca/} provided by the authors. 

\subsection{Experimental Environment}

The experimental environment includes (1) Intel(R) Core(TM) i5-4200U CPU @ 1.60GHz 2.30 GHz 4.0GB RAM laptop running Windows 8.1 (64 bit); (2) MATLAB r2017a.

\section{Experimental Results}\label{sec:experiment_results}

\subsection{RQ1: Does the proposed TOMO outperform traditional over-sampling techniques in CPDP scenario?}

Tables \ref{tb:metricsGM01} and \ref{tb:metricsMCC01} respectively present G-Measure and MCC values in the fashion of 'mean$\pm$std' for our proposed TOMO and SMOTE. The best value on each dataset is in bold. The average row means the average performance across all 34 experiments. The row Win/Tie/Lose shows the results of comparing TOMO with SMOTE(N) based on the Wilcoxon rank-sum test. Win/Tie/Lose represents the number of datasets TOMO wins/ties/loses, compared with the corresponding columns (i.e., SMOTE(100),...,SMOTE(500)). Tables \ref{tb:p-delta_GM01} and \ref{tb:p-delta_MCC01} show the results of P-value and Cliff's delta when comparing TOMO and SMOTE(N) on each dataset in terms of G-Measure and MCC, respectively. 

From Tables \ref{tb:metricsGM01} and \ref{tb:metricsMCC01}, we notice that, \textbf{on average, TOMO improves the G-Measure by 23.7\%$\sim$41.8\%, and the MCC by 54.2\%$\sim$77.8\%, respectively}. This indicates that our proposed TOMO outperforms SMOTE in CPDP scenario. 

More specifically, with respect to G-Measure (Table \ref{tb:metricsGM01}), we notice that TOMO wins 19 datasets, ties 5 datasets, loses 10 datasets, compared with SMOTE(100); TOMO wins 24 datasets, ties 1 dataset, loses 9 datasets compared with SMOTE(200); TOMO wins 22 datasets, ties 3 datasets, loses 9 datasets compared with SMOTE(300); TOMO wins 25 datasets, ties 1 dataset, loses 8 datasets compared with SMOTE(400); TOMO wins 24 datasets, ties 2 datasets, loses 8 datasets compared with SMOTE(400) or SMOTE (500) or SMOTE (500). With respect to MCC (Table \ref{tb:metricsMCC01}), we notice that TOMO wins 21 datasets, ties 5 datasets, loses 8 datasets, compared with SMOTE(100); TOMO wins 24 datasets, ties 2 datasets, loses 8 datasets, compared with SMOTE(200) or SMOTE(300); TOMO wins 24 datasets, ties 3 dataset, loses 7 datasets, compared with SMOTE(400); TOMO wins 25 datasets, ties 2 datasets, loses 7 datasets, compared with SMOTE(500).

%

%


\begin{table}[htbp] %
	\setlength{\abovecaptionskip}{0pt}
	\setlength{\belowcaptionskip}{10pt}
	\caption{\textbf{G-Measure} of Our Transfer Learning Oriented Minority Over-sampling Technique (TOMO) and Traditional Synthetic Minority Over-sampling Technique (SMOTE) \cite{Chawla2002SMOTE}. Both TOMO and SMOTE use the same CPDP approach TNB \cite{Ma2012Transfer}. The best value is in bold.}
	\label{tb:metricsGM01}
	\footnotesize
	\setlength{\leftskip}{-1.8cm}
	\begin{tabular}
		{p{3cm}p{1.8cm}p{1.8cm}p{1.8cm}p{1.8cm}p{1.8cm}p{1.8cm}}
		\hline
		Source$\Rightarrow$Target & TOMO & SMOTE(100) & SMOTE(200) & SMOTE(300) & SMOTE(400) & SMOTE(500) \\ \hline
		
	
		camel-1.0$\Rightarrow$jedit-4.0 & 0.231$\pm$0.039 & 0.642$\pm$0.045 & 0.676$\pm$0.031 & 0.673$\pm$0.023 & \textbf{0.682}$\pm$0.017 & 0.676$\pm$0.015 \\
		ivy-1.4$\Rightarrow$jedit-4.0 & \textbf{0.630}$\pm$0.079 & 0.617$\pm$0.123 & 0.619$\pm$0.124 & 0.601$\pm$0.109 & 0.594$\pm$0.089 & 0.598$\pm$0.090 \\
		tomcat$\Rightarrow$jedit-4.0 & \textbf{0.697}$\pm$0.013 & 0.475$\pm$0.155 & 0.251$\pm$0.140 & 0.169$\pm$0.134 & 0.150$\pm$0.119 & 0.126$\pm$0.096 \\
		prop-6$\Rightarrow$jedit-4.0 & \textbf{0.660}$\pm$0.054 & 0.109$\pm$0.118 & 0.079$\pm$0.079 & 0.075$\pm$0.111 & 0.064$\pm$0.061 & 0.104$\pm$0.088 \\
		poi-2.0$\Rightarrow$jedit-4.0 & \textbf{0.686}$\pm$0.017 & 0.574$\pm$0.083 & 0.565$\pm$0.082 & 0.565$\pm$0.092 & 0.479$\pm$0.108 & 0.394$\pm$0.133 \\
		xalan-2.4$\Rightarrow$jedit-4.0 & 0.681$\pm$0.012 & \textbf{0.685}$\pm$0.014 & 0.661$\pm$0.024 & 0.660$\pm$0.020 & 0.642$\pm$0.035 & 0.613$\pm$0.061 \\
		xerces-1.2$\Rightarrow$jedit-4.0 & \textbf{0.637}$\pm$0.049 & 0.166$\pm$0.167 & 0.082$\pm$0.093 & 0.149$\pm$0.099 & 0.288$\pm$0.142 & 0.360$\pm$0.113 \\
		camel-1.0$\Rightarrow$log4j-1.0 & 0.447$\pm$0.096 & 0.625$\pm$0.113 & 0.691$\pm$0.083 & 0.674$\pm$0.065 & 0.716$\pm$0.026 & \textbf{0.721}$\pm$0.018 \\
		ivy-1.4$\Rightarrow$log4j-1.0 & \textbf{0.732}$\pm$0.051 & 0.505$\pm$0.118 & 0.554$\pm$0.122 & 0.575$\pm$0.126 & 0.567$\pm$0.106 & 0.582$\pm$0.104 \\
		tomcat$\Rightarrow$log4j-1.0 & \textbf{0.677}$\pm$0.060 & 0.300$\pm$0.222 & 0.101$\pm$0.122 & 0.057$\pm$0.094 & 0.052$\pm$0.071 & 0.045$\pm$0.058 \\
		prop-6$\Rightarrow$log4j-1.0 & \textbf{0.714}$\pm$0.051 & 0.165$\pm$0.114 & 0.121$\pm$0.083 & 0.077$\pm$0.089 & 0.090$\pm$0.073 & 0.104$\pm$0.079 \\
		poi-2.0$\Rightarrow$log4j-1.0 & \textbf{0.666}$\pm$0.066 & 0.444$\pm$0.094 & 0.505$\pm$0.148 & 0.553$\pm$0.120 & 0.406$\pm$0.170 & 0.321$\pm$0.168 \\
		xalan-2.4$\Rightarrow$log4j-1.0 & \textbf{0.751}$\pm$0.017 & 0.742$\pm$0.036 & 0.679$\pm$0.070 & 0.662$\pm$0.056 & 0.622$\pm$0.066 & 0.569$\pm$0.124 \\
		xerces-1.2$\Rightarrow$log4j-1.0 & \textbf{0.591}$\pm$0.081 & 0.261$\pm$0.111 & 0.222$\pm$0.094 & 0.313$\pm$0.066 & 0.320$\pm$0.082 & 0.314$\pm$0.102 \\
		camel-1.0$\Rightarrow$ant-1.7 & 0.198$\pm$0.060 & 0.637$\pm$0.073 & 0.685$\pm$0.043 & 0.683$\pm$0.026 & \textbf{0.692}$\pm$0.012 & 0.691$\pm$0.015 \\
		ivy-1.4$\Rightarrow$ant-1.7 & \textbf{0.673}$\pm$0.065 & 0.544$\pm$0.116 & 0.598$\pm$0.093 & 0.605$\pm$0.103 & 0.595$\pm$0.090 & 0.582$\pm$0.091 \\
		tomcat$\Rightarrow$ant-1.7 & \textbf{0.694}$\pm$0.021 & 0.423$\pm$0.192 & 0.185$\pm$0.121 & 0.121$\pm$0.109 & 0.111$\pm$0.097 & 0.093$\pm$0.068 \\
		prop-6$\Rightarrow$ant-1.7 & \textbf{0.627}$\pm$0.049 & 0.241$\pm$0.134 & 0.131$\pm$0.089 & 0.096$\pm$0.115 & 0.103$\pm$0.079 & 0.109$\pm$0.088 \\
		poi-2.0$\Rightarrow$ant-1.7 & \textbf{0.676}$\pm$0.029 & 0.616$\pm$0.050 & 0.580$\pm$0.084 & 0.537$\pm$0.099 & 0.423$\pm$0.122 & 0.342$\pm$0.124 \\
		xalan-2.4$\Rightarrow$ant-1.7 & 0.708$\pm$0.010 & \textbf{0.718}$\pm$0.011 & 0.664$\pm$0.036 & 0.654$\pm$0.036 & 0.617$\pm$0.050 & 0.584$\pm$0.075 \\
		xerces-1.2$\Rightarrow$ant-1.7 & \textbf{0.592}$\pm$0.052 & 0.286$\pm$0.105 & 0.269$\pm$0.102 & 0.351$\pm$0.064 & 0.410$\pm$0.044 & 0.393$\pm$0.066 \\
		camel-1.0$\Rightarrow$xerces-1.2 & 0.180$\pm$0.065 & 0.422$\pm$0.028 & 0.440$\pm$0.028 & 0.436$\pm$0.019 & 0.441$\pm$0.015 & \textbf{0.445}$\pm$0.018 \\
		ivy-1.4$\Rightarrow$xerces-1.2 & 0.430$\pm$0.046 & \textbf{0.530}$\pm$0.113 & 0.521$\pm$0.079 & 0.518$\pm$0.060 & 0.523$\pm$0.049 & 0.526$\pm$0.053 \\
		tomcat$\Rightarrow$xerces-1.2 & \textbf{0.445}$\pm$0.015 & 0.387$\pm$0.121 & 0.275$\pm$0.121 & 0.186$\pm$0.145 & 0.177$\pm$0.147 & 0.152$\pm$0.125 \\
		prop-6$\Rightarrow$xerces-1.2 & \textbf{0.307}$\pm$0.079 & 0.144$\pm$0.125 & 0.073$\pm$0.066 & 0.057$\pm$0.078 & 0.046$\pm$0.046 & 0.069$\pm$0.059 \\
		poi-2.0$\Rightarrow$xerces-1.2 & \textbf{0.488}$\pm$0.028 & 0.464$\pm$0.026 & 0.470$\pm$0.036 & 0.478$\pm$0.025 & 0.427$\pm$0.098 & 0.354$\pm$0.142 \\
		xalan-2.4$\Rightarrow$xerces-1.2 & 0.457$\pm$0.017 & 0.464$\pm$0.012 & 0.490$\pm$0.017 & 0.494$\pm$0.024 & \textbf{0.500}$\pm$0.020 & 0.476$\pm$0.055 \\
		camel-1.0$\Rightarrow$velocity-1.6 & 0.225$\pm$0.041 & 0.380$\pm$0.100 & 0.469$\pm$0.075 & 0.452$\pm$0.085 & \textbf{0.509}$\pm$0.035 & 0.500$\pm$0.048 \\
		ivy-1.4$\Rightarrow$velocity-1.6 & 0.527$\pm$0.069 & 0.407$\pm$0.106 & 0.448$\pm$0.125 & 0.500$\pm$0.136 & 0.514$\pm$0.110 & \textbf{0.530}$\pm$0.091 \\
		tomcat$\Rightarrow$velocity-1.6 & \textbf{0.497}$\pm$0.053 & 0.413$\pm$0.208 & 0.215$\pm$0.153 & 0.148$\pm$0.140 & 0.143$\pm$0.133 & 0.111$\pm$0.093 \\
		prop-6$\Rightarrow$velocity-1.6 & \textbf{0.398}$\pm$0.099 & 0.137$\pm$0.138 & 0.078$\pm$0.094 & 0.070$\pm$0.123 & 0.062$\pm$0.073 & 0.121$\pm$0.110 \\
		poi-2.0$\Rightarrow$velocity-1.6 & 0.523$\pm$0.049 & \textbf{0.576}$\pm$0.074 & 0.571$\pm$0.080 & 0.553$\pm$0.095 & 0.443$\pm$0.135 & 0.361$\pm$0.148 \\
		xalan-2.4$\Rightarrow$velocity-1.6 & 0.508$\pm$0.058 & 0.575$\pm$0.046 & 0.637$\pm$0.022 & \textbf{0.639}$\pm$0.020 & 0.627$\pm$0.020 & 0.603$\pm$0.070 \\
		xerces-1.2$\Rightarrow$velocity-1.6 & \textbf{0.496}$\pm$0.072 & 0.259$\pm$0.154 & 0.231$\pm$0.129 & 0.343$\pm$0.111 & 0.453$\pm$0.082 & 0.463$\pm$0.047 \\
		\textbf{Win/Tie/Lose} & ---  &  19/5/10 &  24/1/9 &  22/3/9 &  25/1/8 &  24/2/8 \\
		\textbf{Average} & 0.543$\pm$0.166 & 0.439$\pm$0.181 & 0.407$\pm$0.221 & 0.404$\pm$0.227 & 0.397$\pm$0.219 & 0.383$\pm$0.211 \\
		
		\hline
	\end{tabular}
\end{table}

\begin{table}[htbp] %
	\setlength{\abovecaptionskip}{0pt}
	\setlength{\belowcaptionskip}{10pt}
	\caption{\textbf{MCC} of Our Transfer Learning Oriented Minority Over-sampling Technique (TOMO) and Traditional Synthetic Minority Over-sampling Technique SMOTE \cite{Chawla2002SMOTE}. Both TOMO and SMOTE use the same CPDP approach TNB \cite{Ma2012Transfer}. The best value is in bold.}
	\label{tb:metricsMCC01}
	\footnotesize
	\setlength{\leftskip}{-1.8cm}
	\begin{tabular}
		{p{3cm}p{1.8cm}p{1.8cm}p{1.8cm}p{1.8cm}p{1.8cm}p{1.8cm}}
		\hline
		Source$\Rightarrow$Target & TOMO & SMOTE(100) & SMOTE(200) & SMOTE(300) & SMOTE(400) & SMOTE(500) \\ \hline	

		camel-1.0$\Rightarrow$jedit-4.0 & 0.107$\pm$0.034 & 0.328$\pm$0.040 & \textbf{0.338}$\pm$0.025 & 0.333$\pm$0.027 & 0.326$\pm$0.031 & 0.316$\pm$0.023 \\
		ivy-1.4$\Rightarrow$jedit-4.0 & \textbf{0.318}$\pm$0.073 & 0.266$\pm$0.073 & 0.280$\pm$0.079 & 0.266$\pm$0.064 & 0.257$\pm$0.050 & 0.262$\pm$0.059 \\
		tomcat$\Rightarrow$jedit-4.0 & \textbf{0.452}$\pm$0.022 & 0.236$\pm$0.072 & 0.166$\pm$0.059 & 0.126$\pm$0.069 & 0.115$\pm$0.053 & 0.103$\pm$0.052 \\
		prop-6$\Rightarrow$jedit-4.0 & \textbf{0.374}$\pm$0.055 & 0.038$\pm$0.086 & 0.063$\pm$0.071 & 0.058$\pm$0.076 & 0.068$\pm$0.054 & 0.086$\pm$0.061 \\
		poi-2.0$\Rightarrow$jedit-4.0 & \textbf{0.341}$\pm$0.031 & 0.274$\pm$0.055 & 0.272$\pm$0.043 & 0.279$\pm$0.050 & 0.247$\pm$0.043 & 0.224$\pm$0.062 \\
		xalan-2.4$\Rightarrow$jedit-4.0 & \textbf{0.387}$\pm$0.020 & 0.339$\pm$0.036 & 0.309$\pm$0.029 & 0.306$\pm$0.026 & 0.308$\pm$0.029 & 0.296$\pm$0.043 \\
		xerces-1.2$\Rightarrow$jedit-4.0 & \textbf{0.258}$\pm$0.055 & 0.109$\pm$0.082 & 0.070$\pm$0.058 & 0.096$\pm$0.053 & 0.125$\pm$0.064 & 0.132$\pm$0.044 \\
		camel-1.0$\Rightarrow$log4j-1.0 & 0.381$\pm$0.079 & \textbf{0.488}$\pm$0.065 & 0.482$\pm$0.058 & 0.477$\pm$0.034 & 0.471$\pm$0.041 & 0.475$\pm$0.044 \\
		ivy-1.4$\Rightarrow$log4j-1.0 & \textbf{0.495}$\pm$0.068 & 0.110$\pm$0.099 & 0.224$\pm$0.102 & 0.261$\pm$0.071 & 0.230$\pm$0.080 & 0.271$\pm$0.076 \\
		tomcat$\Rightarrow$log4j-1.0 & \textbf{0.456}$\pm$0.068 & 0.202$\pm$0.107 & 0.099$\pm$0.073 & 0.062$\pm$0.069 & 0.062$\pm$0.062 & 0.055$\pm$0.054 \\
		prop-6$\Rightarrow$log4j-1.0 & \textbf{0.443}$\pm$0.059 & 0.128$\pm$0.058 & 0.105$\pm$0.073 & 0.053$\pm$0.065 & 0.048$\pm$0.059 & 0.064$\pm$0.077 \\
		poi-2.0$\Rightarrow$log4j-1.0 & \textbf{0.371}$\pm$0.094 & 0.147$\pm$0.071 & 0.205$\pm$0.119 & 0.261$\pm$0.081 & 0.190$\pm$0.086 & 0.190$\pm$0.083 \\
		xalan-2.4$\Rightarrow$log4j-1.0 & \textbf{0.476}$\pm$0.036 & 0.452$\pm$0.048 & 0.404$\pm$0.059 & 0.397$\pm$0.044 & 0.374$\pm$0.044 & 0.354$\pm$0.066 \\
		xerces-1.2$\Rightarrow$log4j-1.0 & \textbf{0.223}$\pm$0.128 & -0.153$\pm$0.066 & -0.165$\pm$0.076 & -0.187$\pm$0.064 & -0.197$\pm$0.059 & -0.199$\pm$0.070 \\
		camel-1.0$\Rightarrow$ant-1.7 & 0.021$\pm$0.031 & 0.347$\pm$0.020 & 0.348$\pm$0.020 & \textbf{0.350}$\pm$0.021 & 0.336$\pm$0.015 & 0.336$\pm$0.014 \\
		ivy-1.4$\Rightarrow$ant-1.7 & \textbf{0.334}$\pm$0.073 & 0.183$\pm$0.104 & 0.247$\pm$0.073 & 0.257$\pm$0.076 & 0.250$\pm$0.065 & 0.249$\pm$0.047 \\
		tomcat$\Rightarrow$ant-1.7 & \textbf{0.383}$\pm$0.023 & 0.218$\pm$0.096 & 0.128$\pm$0.052 & 0.104$\pm$0.048 & 0.094$\pm$0.057 & 0.091$\pm$0.035 \\
		prop-6$\Rightarrow$ant-1.7 & \textbf{0.263}$\pm$0.064 & 0.057$\pm$0.070 & 0.053$\pm$0.035 & 0.040$\pm$0.048 & 0.042$\pm$0.040 & 0.032$\pm$0.047 \\
		poi-2.0$\Rightarrow$ant-1.7 & \textbf{0.311}$\pm$0.029 & 0.260$\pm$0.035 & 0.255$\pm$0.054 & 0.237$\pm$0.053 & 0.191$\pm$0.056 & 0.163$\pm$0.055 \\
		xalan-2.4$\Rightarrow$ant-1.7 & 0.365$\pm$0.015 & \textbf{0.381}$\pm$0.017 & 0.340$\pm$0.029 & 0.332$\pm$0.030 & 0.309$\pm$0.038 & 0.288$\pm$0.041 \\
		xerces-1.2$\Rightarrow$ant-1.7 & \textbf{0.179}$\pm$0.090 & -0.09$\pm$0.056 & -0.10$\pm$0.061 & -0.127$\pm$0.051 & -0.095$\pm$0.055 & -0.111$\pm$0.065 \\
		camel-1.0$\Rightarrow$xerces-1.2 & 0.042$\pm$0.039 & 0.100$\pm$0.064 & 0.086$\pm$0.056 & \textbf{0.102}$\pm$0.060 & 0.060$\pm$0.051 & 0.071$\pm$0.044 \\
		ivy-1.4$\Rightarrow$xerces-1.2 & 0.088$\pm$0.025 & \textbf{0.150}$\pm$0.094 & 0.139$\pm$0.041 & 0.121$\pm$0.043 & 0.125$\pm$0.052 & 0.120$\pm$0.047 \\
		tomcat$\Rightarrow$xerces-1.2 & \textbf{0.125}$\pm$0.020 & -0.097$\pm$0.087 & -0.146$\pm$0.071 & -0.124$\pm$0.087 & -0.105$\pm$0.084 & -0.107$\pm$0.084 \\
		prop-6$\Rightarrow$xerces-1.2 & \textbf{0.039}$\pm$0.033 & 0.018$\pm$0.053 & 0.026$\pm$0.050 & 0.022$\pm$0.040 & 0.021$\pm$0.041 & 0.034$\pm$0.036 \\
		poi-2.0$\Rightarrow$xerces-1.2 & \textbf{0.141}$\pm$0.027 & 0.013$\pm$0.041 & 0.030$\pm$0.036 & 0.037$\pm$0.035 & 0.019$\pm$0.048 & -0.003$\pm$0.062 \\
		xalan-2.4$\Rightarrow$xerces-1.2 & \textbf{0.107}$\pm$0.019 & 0.076$\pm$0.034 & 0.028$\pm$0.032 & 0.029$\pm$0.033 & 0.031$\pm$0.036 & 0.017$\pm$0.047 \\
		camel-1.0$\Rightarrow$velocity-1.6 & 0.153$\pm$0.039 & 0.188$\pm$0.042 & 0.218$\pm$0.037 & 0.217$\pm$0.033 & \textbf{0.221}$\pm$0.040 & 0.209$\pm$0.036 \\
		ivy-1.4$\Rightarrow$velocity-1.6 & \textbf{0.252}$\pm$0.067 & 0.043$\pm$0.054 & 0.107$\pm$0.082 & 0.134$\pm$0.060 & 0.155$\pm$0.079 & 0.178$\pm$0.065 \\
		tomcat$\Rightarrow$velocity-1.6 & \textbf{0.238}$\pm$0.035 & 0.228$\pm$0.079 & 0.166$\pm$0.063 & 0.128$\pm$0.063 & 0.124$\pm$0.083 & 0.114$\pm$0.066 \\
		prop-6$\Rightarrow$velocity-1.6 & \textbf{0.145}$\pm$0.068 & 0.093$\pm$0.068 & 0.061$\pm$0.082 & 0.046$\pm$0.080 & 0.060$\pm$0.073 & 0.090$\pm$0.079 \\
		poi-2.0$\Rightarrow$velocity-1.6 & 0.187$\pm$0.050 & 0.213$\pm$0.063 & 0.234$\pm$0.053 & \textbf{0.237}$\pm$0.053 & 0.228$\pm$0.043 & 0.204$\pm$0.058 \\
		xalan-2.4$\Rightarrow$velocity-1.6 & 0.183$\pm$0.041 & 0.243$\pm$0.033 & \textbf{0.281}$\pm$0.027 & 0.279$\pm$0.031 & 0.269$\pm$0.032 & 0.273$\pm$0.040 \\
		xerces-1.2$\Rightarrow$velocity-1.6 & 0.079$\pm$0.072 & 0.045$\pm$0.069 & 0.097$\pm$0.052 & \textbf{0.111}$\pm$0.057 & 0.066$\pm$0.067 & 0.005$\pm$0.080 \\
		\textbf{Win/Tie/Lose} & ---  &  21/5/8 &  24/2/8 &  24/2/8 &  24/3/7 &  25/2/7 \\
		\textbf{Average} & 0.256$\pm$0.140 & 0.166$\pm$0.15 & 0.160$\pm$0.149 & 0.156$\pm$0.153 & 0.148$\pm$0.145 & 0.144$\pm$0.146 \\
		
		\hline
	\end{tabular}
\end{table}


%
%

The reason that the proposed TOMO can achieve such good performance is that TOMO generates artificial minority class samples by taking the distribution of source data and the distribution of target data into consideration. This can not only balance the training data but reduce the distribution differences between the source and target datasets. However, we also notice that in a few cases TOMO achieves poor performance. For example, for camel-1.0$\Rightarrow$jedit-4.0, TOMO performs badly in terms of both G-Measure and MCC compared with baselines. Possible reasons include: (1) For our TOMO, there is a hyper-parameter $\lambda$, which is used to adjust the weights of $distSS_N$ and $distTS_N$ (Step-11 in Algorithm \ref{alg:TOMO}) when constructing nearest-neighbor matrix. $\lambda$ has an effect on the performance of TOMO. However, in this paper, for TOMO, we just set $\lambda$ to 0.4. (2). For TOMO, we just use k-means \cite{Jain2008Data} to find potential minority class samples in target dataset. The performance of k-means will affect the performance of our proposed TOMO approach.

Tables \ref{tb:p-delta_GM01} and \ref{tb:p-delta_MCC01} show the results of P-value and Cliff's delta when comparing TOMO and SMOTE on each dataset in terms of G-Measure and MCC, respectively.


\begin{table}[htbp] %
	\setlength{\abovecaptionskip}{0pt}
	\setlength{\belowcaptionskip}{10pt}
	\caption{\textbf{P-value} and \textbf{Cliff's delta} When Comparing our Proposed TOMO with Traditional Minority Over-sampling Technique SMOTE in terms of \textbf{G-Measure}}
	\label{tb:p-delta_GM01}
	\centering
	\footnotesize
	\setlength{\leftskip}{-2.6cm}
	\begin{tabular}{lllllllllll}
		\hline
		\multirow{2}{*}{Source$\Rightarrow$Target} & \multicolumn{2}{c}{TOMO vs SMOTE(100)} & \multicolumn{2}{c}{TOMO vs SMOTE(200)} & \multicolumn{2}{c}{TOMO vs SMOTE(300)} & \multicolumn{2}{c}{TOMO vs SMOTE(400)} & \multicolumn{2}{c}{TOMO vs SMOTE(500)} \\
		& P-Value & $\delta$ & P-Value & $\delta$ & P-Value & $\delta$ & P-Value & $\delta$ & P-Value & $\delta$ \\ \hline
		camel-1.0$\Rightarrow$jedit-4.0 & 3.01e-11 & -1 & 3.00e-11 & -1 & 3.01e-11 & -1 & 3.01e-11 & -1 & 3.01e-11 & -1 \\
		ivy-1.4$\Rightarrow$jedit-4.0 & 0.249 & 0.174 & 0.965 & -0.008 & 0.092 & 0.254 & 0.013 & 0.376 & 0.054 & 0.291 \\
		tomcat$\Rightarrow$jedit-4.0 & 1.45e-10 & 0.964 & 3.00e-11 & 1 & 2.99e-11 & 1 & 2.97e-11 & 1 & 2.99e-11 & 1 \\
		prop-6$\Rightarrow$jedit-4.0 & 2.97e-11 & 1 & 2.97e-11 & 1 & 2.52e-11 & 1 & 2.93e-11 & 1 & 2.95e-11 & 1 \\
		poi-2.0$\Rightarrow$jedit-4.0 & 3.82e-10 & 0.942 & 6.69e-11 & 0.982 & 4.97e-11 & 0.989 & 3.02e-11 & 1 & 3.01e-11 & 1 \\
		xalan-2.4$\Rightarrow$jedit-4.0 & 0.151 & -0.217 & 2.83e-4 & 0.547 & 2.76e-5 & 0.631 & 5.98e-8 & 0.816 & 6.47e-9 & 0.873 \\
		xerces-1.2$\Rightarrow$jedit-4.0 & 4.96e-11 & 0.989 & 2.94e-11 & 1 & 3.01e-11 & 1 & 3.69e-11 & 0.996 & 9.92e-11 & 0.973 \\
		camel-1.0$\Rightarrow$log4j-1.0 & 5.43e-8 & -0.818 & 7.25e-10 & -0.927 & 6.48e-11 & -0.982 & 2.95e-11 & -1 & 2.96e-11 & -1 \\
		ivy-1.4$\Rightarrow$log4j-1.0 & 9.84e-11 & 0.973 & 1.06e-9 & 0.918 & 3.32e-9 & 0.890 & 1.17e-9 & 0.916 & 5.93e-9 & 0.876 \\
		tomcat$\Rightarrow$log4j-1.0 & 9.23e-9 & 0.864 & 2.86e-11 & 1 & 2.51e-11 & 1 & 2.54e-11 & 1 & 2.47e-11 & 1 \\
		prop-6$\Rightarrow$log4j-1.0 & 2.97e-11 & 1 & 2.94e-11 & 1 & 2.75e-11 & 1 & 2.90e-11 & 1 & 2.88e-11 & 1 \\
		poi-2.0$\Rightarrow$log4j-1.0 & 7.32e-10 & 0.927 & 1.63e-5 & 0.649 & 6.55e-5 & 0.601 & 3.07e-8 & 0.833 & 7.35e-10 & 0.927 \\
		xalan-2.4$\Rightarrow$log4j-1.0 & 0.801 & 0.039 & 7.63e-5 & 0.596 & 1.30e-8 & 0.856 & 8.10e-11 & 0.978 & 5.94e-9 & 0.876 \\
		xerces-1.2$\Rightarrow$log4j-1.0 & 4.96e-11 & 0.989 & 4.07e-11 & 0.993 & 4.08e-11 & 0.993 & 3.69e-11 & 0.996 & 4.50e-11 & 0.991 \\
		camel-1.0$\Rightarrow$ant-1.7 & 3.02e-11 & -1 & 3.02e-11 & -1 & 3.02e-11 & -1 & 3.02e-11 & -1 & 3.02e-11 & -1 \\
		ivy-1.4$\Rightarrow$ant-1.7 & 7.09e-8 & 0.811 & 2.15e-6 & 0.713 & 3.83e-5 & 0.620 & 2.88e-6 & 0.704 & 8.84e-7 & 0.740 \\
		tomcat$\Rightarrow$ant-1.7 & 4.57e-9 & 0.882 & 3.02e-11 & 1 & 3.01e-11 & 1 & 3.02e-11 & 1 & 3.01e-11 & 1 \\
		prop-6$\Rightarrow$ant-1.7 & 3.02e-11 & 1 & 3.02e-11 & 1 & 2.91e-11 & 1 & 3.01e-11 & 1 & 2.98e-11 & 1 \\
		poi-2.0$\Rightarrow$ant-1.7 & 5.00e-9 & 0.880 & 1.69e-9 & 0.907 & 4.20e-10 & 0.940 & 4.97e-11 & 0.989 & 3.02e-11 & 1 \\
		xalan-2.4$\Rightarrow$ant-1.7 & 0.002 & -0.456 & 1.87e-7 & 0.784 & 9.26e-9 & 0.864 & 1.77e-10 & 0.960 & 7.38e-11 & 0.980 \\
		xerces-1.2$\Rightarrow$ant-1.7 & 3.01e-11 & 1 & 3.02e-11 & 1 & 3.02e-11 & 1 & 3.02e-11 & 1 & 3.02e-11 & 1 \\
		camel-1.0$\Rightarrow$xerces-1.2 & 3.01e-11 & -1 & 3.01e-11 & -1 & 3.01e-11 & -1 & 3.01e-11 & -1 & 2.98e-11 & -1 \\
		ivy-1.4$\Rightarrow$xerces-1.2 & 1.30e-8 & -0.856 & 9.41e-8 & -0.803 & 2.59e-8 & -0.838 & 1.42e-8 & -0.853 & 3.48e-9 & -0.889 \\
		tomcat$\Rightarrow$xerces-1.2 & 0.318 & 0.151 & 3.34e-8 & 0.831 & 2.03e-9 & 0.902 & 5.01e-10 & 0.936 & 3.00e-11 & 1 \\
		prop-6$\Rightarrow$xerces-1.2 & 6.27e-6 & 0.680 & 5.53e-10 & 0.933 & 3.30e-10 & 0.944 & 8.01e-11 & 0.978 & 1.91e-10 & 0.958 \\
		poi-2.0$\Rightarrow$xerces-1.2 & 0.002 & 0.478 & 0.025 & 0.338 & 0.096 & 0.251 & 0.009 & 0.393 & 5.07e-6 & 0.687 \\
		xalan-2.4$\Rightarrow$xerces-1.2 & 0.038 & -0.313 & 2.83e-8 & -0.836 & 3.95e-8 & -0.827 & 9.24e-9 & -0.864 & 3.37e-4 & -0.54 \\
		camel-1.0$\Rightarrow$velocity-1.6 & 7.65e-10 & -0.926 & 2.99e-11 & -1 & 3.00e-11 & -1 & 3.00e-11 & -1 & 3.00e-11 & -1 \\
		ivy-1.4$\Rightarrow$velocity-1.6 & 4.41e-6 & 0.691 & 0.004 & 0.436 & 0.865 & -0.027 & 0.796 & -0.04 & 0.483 & -0.107 \\
		tomcat$\Rightarrow$velocity-1.6 & 0.297 & 0.158 & 4.28e-8 & 0.824 & 1.05e-9 & 0.918 & 8.00e-11 & 0.978 & 3.27e-11 & 0.998 \\
		prop-6$\Rightarrow$velocity-1.6 & 2.81e-8 & 0.836 & 2.04e-10 & 0.956 & 7.39e-10 & 0.918 & 9.17e-11 & 0.973 & 8.48e-10 & 0.922 \\
		poi-2.0$\Rightarrow$velocity-1.6 & 5.97e-5 & -0.604 & 3.99e-4 & -0.533 & 0.022 & -0.344 & 0.03 & 0.327 & 6.28e-6 & 0.680 \\
		xalan-2.4$\Rightarrow$velocity-1.6 & 2.95e-5 & -0.629 & 1.77e-10 & -0.96 & 2.15e-10 & -0.956 & 6.39e-10 & -0.93 & 1.19e-6 & -0.731 \\
		xerces-1.2$\Rightarrow$velocity-1.6 & 1.84e-8 & 0.847 & 2.36e-10 & 0.953 & 9.06e-8 & 0.804 & 0.016 & 0.362 & 0.012 & 0.378 \\
		
		\hline
	\end{tabular}	
\end{table}

\begin{table}[htbp] %
	\setlength{\abovecaptionskip}{0pt}
	\setlength{\belowcaptionskip}{10pt}
	\caption{ \textbf{P-value} and \textbf{Cliff's delta} When Comparing our Proposed TOMO with Traditional Minority Over-sampling Technique SMOTE in terms of \textbf{MCC}}
	\label{tb:p-delta_MCC01}
	\centering
	\footnotesize
	\setlength{\leftskip}{-2.6cm}
	\begin{tabular}{lllllllllll}
		\hline
		\multirow{2}{*}{Source$\Rightarrow$Target} & \multicolumn{2}{c}{TOMO vs SMOTE(100)} & \multicolumn{2}{c}{TOMO vs SMOTE(200)} & \multicolumn{2}{c}{TOMO vs SMOTE(300)} & \multicolumn{2}{c}{TOMO vs SMOTE(400)} & \multicolumn{2}{c}{TOMO vs SMOTE(500)} \\
		& P-Value & $\delta$ & P-Value & $\delta$ & P-Value & $\delta$ & P-Value & $\delta$ & P-Value & $\delta$ \\ \hline
		
		camel-1.0$\Rightarrow$jedit-4.0 & 3.01e-11 & -1 & 3.00e-11 & -1 & 3.01e-11 & -1 & 3.01e-11 & -1 & 3.01e-11 & -1 \\
		ivy-1.4$\Rightarrow$jedit-4.0 & 4.58e-4 & 0.528 & 0.009 & 0.394 & 2.46e-4 & 0.552 & 2.28e-5 & 0.638 & 2.53e-4 & 0.551 \\
		tomcat$\Rightarrow$jedit-4.0 & 3.00e-11 & 1 & 3.00e-11 & 1 & 2.99e-11 & 1 & 2.97e-11 & 1 & 2.99e-11 & 1 \\
		prop-6$\Rightarrow$jedit-4.0 & 2.97e-11 & 1 & 2.99e-11 & 1 & 2.52e-11 & 1 & 2.93e-11 & 1 & 2.95e-11 & 1 \\
		poi-2.0$\Rightarrow$jedit-4.0 & 2.03e-7 & 0.782 & 2.39e-8 & 0.840 & 4.11e-7 & 0.762 & 2.61e-10 & 0.951 & 3.47e-10 & 0.944 \\
		xalan-2.4$\Rightarrow$jedit-4.0 & 6.24e-7 & 0.750 & 4.93e-11 & 0.989 & 2.99e-11 & 1 & 2.14e-10 & 0.956 & 4.47e-11 & 0.991 \\
		xerces-1.2$\Rightarrow$jedit-4.0 & 1.85e-8 & 0.847 & 1.29e-10 & 0.967 & 1.32e-10 & 0.967 & 3.82e-9 & 0.887 & 1.17e-9 & 0.916 \\
		camel-1.0$\Rightarrow$log4j-1.0 & 4.05e-7 & -0.762 & 6.45e-7 & -0.749 & 8.17e-8 & -0.807 & 6.96e-7 & -0.747 & 1.18e-6 & -0.731 \\
		ivy-1.4$\Rightarrow$log4j-1.0 & 3.31e-11 & 0.998 & 8.09e-11 & 0.978 & 1.03e-10 & 0.972 & 7.33e-11 & 0.980 & 1.20e-10 & 0.969 \\
		tomcat$\Rightarrow$log4j-1.0 & 2.36e-10 & 0.953 & 2.86e-11 & 1 & 2.51e-11 & 1 & 2.54e-11 & 1 & 2.47e-11 & 1 \\
		prop-6$\Rightarrow$log4j-1.0 & 2.97e-11 & 1 & 2.94e-11 & 1 & 2.75e-11 & 1 & 2.90e-11 & 1 & 2.88e-11 & 1 \\
		poi-2.0$\Rightarrow$log4j-1.0 & 9.68e-10 & 0.920 & 3.32e-6 & 0.700 & 3.47e-5 & 0.623 & 3.64e-8 & 0.829 & 5.52e-8 & 0.818 \\
		xalan-2.4$\Rightarrow$log4j-1.0 & 0.071 & 0.272 & 9.48e-6 & 0.667 & 3.63e-8 & 0.829 & 5.54e-10 & 0.933 & 3.81e-9 & 0.887 \\
		xerces-1.2$\Rightarrow$log4j-1.0 & 3.33e-11 & 0.998 & 3.69e-11 & 0.996 & 3.02e-11 & 1 & 3.02e-11 & 1 & 3.02e-11 & 1 \\
		camel-1.0$\Rightarrow$ant-1.7 & 3.02e-11 & -1 & 3.02e-11 & -1 & 3.02e-11 & -1 & 3.02e-11 & -1 & 3.02e-11 & -1 \\
		ivy-1.4$\Rightarrow$ant-1.7 & 1.47e-7 & 0.791 & 4.12e-6 & 0.693 & 8.29e-6 & 0.671 & 2.49e-6 & 0.709 & 1.11e-6 & 0.733 \\
		tomcat$\Rightarrow$ant-1.7 & 6.72e-10 & 0.929 & 3.02e-11 & 1 & 3.01e-11 & 1 & 3.02e-11 & 1 & 3.01e-11 & 1 \\
		prop-6$\Rightarrow$ant-1.7 & 6.70e-11 & 0.982 & 3.34e-11 & 0.998 & 3.22e-11 & 0.998 & 3.01e-11 & 1 & 2.98e-11 & 1 \\
		poi-2.0$\Rightarrow$ant-1.7 & 3.25e-7 & 0.769 & 2.15e-6 & 0.713 & 9.83e-8 & 0.802 & 1.96e-10 & 0.958 & 4.50e-11 & 0.991 \\
		xalan-2.4$\Rightarrow$ant-1.7 & 3.37e-4 & -0.54 & 3.56e-4 & 0.538 & 3.32e-6 & 0.700 & 1.70e-8 & 0.849 & 1.29e-9 & 0.913 \\
		xerces-1.2$\Rightarrow$ant-1.7 & 3.68e-11 & 0.996 & 3.69e-11 & 0.996 & 3.02e-11 & 1 & 3.02e-11 & 1 & 3.34e-11 & 0.998 \\
		camel-1.0$\Rightarrow$xerces-1.2 & 8.14e-5 & -0.593 & 0.001 & -0.482 & 1.32e-4 & -0.576 & 0.099 & -0.249 & 0.002 & -0.467 \\
		ivy-1.4$\Rightarrow$xerces-1.2 & 3.56e-6 & -0.698 & 6.76e-7 & -0.748 & 0.008 & -0.402 & 0.002 & -0.476 & 0.013 & -0.373 \\
		tomcat$\Rightarrow$xerces-1.2 & 3.01e-11 & 1 & 3.00e-11 & 1 & 3.00e-11 & 1 & 2.98e-11 & 1 & 3.00e-11 & 1 \\
		prop-6$\Rightarrow$xerces-1.2 & 0.122 & 0.233 & 0.631 & 0.073 & 0.079 & 0.264 & 0.093 & 0.253 & 0.446 & 0.116 \\
		poi-2.0$\Rightarrow$xerces-1.2 & 6.63e-11 & 0.982 & 8.92e-11 & 0.976 & 7.33e-11 & 0.980 & 7.33e-11 & 0.980 & 8.92e-11 & 0.976 \\
		xalan-2.4$\Rightarrow$xerces-1.2 & 2.00e-4 & 0.560 & 1.20e-10 & 0.969 & 1.32e-10 & 0.967 & 3.47e-10 & 0.944 & 6.68e-11 & 0.982 \\
		camel-1.0$\Rightarrow$velocity-1.6 & 7.88e-4 & -0.506 & 1.15e-7 & -0.798 & 9.02e-8 & -0.804 & 1.59e-7 & -0.789 & 1.60e-6 & -0.722 \\
		ivy-1.4$\Rightarrow$velocity-1.6 & 4.95e-11 & 0.989 & 9.05e-8 & 0.804 & 9.82e-8 & 0.802 & 1.87e-5 & 0.644 & 2.00e-4 & 0.560 \\
		tomcat$\Rightarrow$velocity-1.6 & 0.589 & -0.082 & 5.83e-6 & 0.682 & 4.24e-8 & 0.824 & 5.54e-7 & 0.753 & 1.52e-9 & 0.909 \\
		prop-6$\Rightarrow$velocity-1.6 & 0.005 & 0.427 & 1.47e-4 & 0.571 & 9.16e-5 & 0.584 & 7.04e-5 & 0.598 & 0.012 & 0.378 \\
		poi-2.0$\Rightarrow$velocity-1.6 & 0.119 & -0.236 & 0.001 & -0.496 & 5.87e-4 & -0.518 & 0.002 & -0.46 & 0.246 & -0.176 \\
		xalan-2.4$\Rightarrow$velocity-1.6 & 4.10e-7 & -0.762 & 9.73e-10 & -0.92 & 9.73e-10 & -0.92 & 5.21e-9 & -0.879 & 2.66e-9 & -0.896 \\
		xerces-1.2$\Rightarrow$velocity-1.6 & 0.082 & 0.262 & 0.464 & -0.111 & 0.066 & -0.278 & 0.412 & 0.124 & 4.98e-4 & 0.524 \\
		
		\hline
	\end{tabular}	
\end{table}

\subsection{RQ2: Is feature weighting strategy helpful for TNB?}

Table \ref{tb:metricsGM_MCC02} shows G-Measure and MCC values in the fashion of 'mean$\pm$std' for our proposed FWTNB and the baseline (i.e., TNB) in CPDP scenario. The best value on each dataset is in bold. The average row means the average performance across all 34 experiments. The row Win/Tie/Lose shows the results of comparing FWTNB with TNB based on the Wilcoxon rank-sum test. Win/Tie/Lose represents the number of datasets FWTNB wins/tie/loses, compared with TNB. 

From Table \ref{tb:metricsGM_MCC02}, we notice that for G-Measure, FWTNB wins 15 datasets, ties 9 datasets, and loses 10 datasets, compared with TNB in terms of G-Measure; for MCC, FWTNB wins 16 datasets, ties 8 datasets, and loses 10 datasets compared TNB. We also notice that \textbf{FWTNB improves the average G-Measure by 11\%, and the average MCC by 29.2\%, compared with TNB}. This indicates our feature weighting strategy is helpful for improving the performance of TNB.

This is because FWTNB pays different attention to the features with different importance. Features that are more relevant with software defect should play a more critical role in predicting the label of a software module, otherwise, play a less important role. Additionally, FWTNB ties even lose in some cases (e.g., camel-1.0$\Rightarrow$log4j-1.0 ), compared with TNB. Possible reasons include: (1) features have the similar or the same importance. (2) FWTNB has an important hyper-parameter $\sigma$, which has an effect on the performance of FWTNB. For different datasets, the optimal value of $\sigma$ may be different. However, in this study, we just set $\sigma$ to 1. 


\begin{table}[htbp] %
	\setlength{\abovecaptionskip}{0pt}
	\setlength{\belowcaptionskip}{5pt}
	\caption{\textbf{G-Measure} and \textbf{MCC} of Our proposed FWTNB and the baseline. The best value is in bold.}
	\label{tb:metricsGM_MCC02}
	\footnotesize
	\centering
	\begin{tabular}{lllll}
		\hline
		\multirow{2}{*}{Source$\Rightarrow$Target} & \multicolumn{2}{c}{G-Measure} & \multicolumn{2}{c}{MCC} \\  
		& FWTNB & TNB & FWTNB & TNB \\ \hline 
		camel-1.0$\Rightarrow$jedit-4.0 & 0.625$\pm$0.058 & \textbf{0.652}$\pm$0.037 & \textbf{0.333}$\pm$0.037 & 0.322$\pm$0.034 \\
		ivy-1.4$\Rightarrow$jedit-4.0 & 0.591$\pm$0.064 & \textbf{0.606}$\pm$0.126 & \textbf{0.284}$\pm$0.065 & 0.256$\pm$0.072 \\
		tomcat$\Rightarrow$jedit-4.0 & \textbf{0.566}$\pm$0.119 & 0.431$\pm$0.164 & \textbf{0.294}$\pm$0.064 & 0.223$\pm$0.070 \\
		prop-6$\Rightarrow$jedit-4.0 & \textbf{0.296}$\pm$0.115 & 0.105$\pm$0.077 & \textbf{0.188}$\pm$0.046 & 0.050$\pm$0.063 \\
		poi-2.0$\Rightarrow$jedit-4.0 & \textbf{0.675}$\pm$0.027 & 0.584$\pm$0.084 & \textbf{0.321}$\pm$0.031 & 0.278$\pm$0.044 \\
		xalan-2.4$\Rightarrow$jedit-4.0 & \textbf{0.685}$\pm$0.010 & 0.680$\pm$0.015 & \textbf{0.357}$\pm$0.030 & 0.328$\pm$0.034 \\
		xerces-1.2$\Rightarrow$jedit-4.0 & \textbf{0.399}$\pm$0.108 & 0.073$\pm$0.056 & \textbf{0.217}$\pm$0.043 & 0.071$\pm$0.036 \\
		camel-1.0$\Rightarrow$log4j-1.0 & 0.526$\pm$0.171 & \textbf{0.653}$\pm$0.088 & 0.451$\pm$0.120 & \textbf{0.479}$\pm$0.059 \\
		ivy-1.4$\Rightarrow$log4j-1.0 & 0.380$\pm$0.140 & \textbf{0.479}$\pm$0.126 & \textbf{0.103}$\pm$0.097 & 0.094$\pm$0.123 \\
		tomcat$\Rightarrow$log4j-1.0 & \textbf{0.614}$\pm$0.156 & 0.250$\pm$0.204 & \textbf{0.368}$\pm$0.111 & 0.177$\pm$0.095 \\
		prop-6$\Rightarrow$log4j-1.0 & \textbf{0.199}$\pm$0.130 & 0.162$\pm$0.073 & 0.099$\pm$0.085 & \textbf{0.131}$\pm$0.055 \\
		poi-2.0$\Rightarrow$log4j-1.0 & \textbf{0.667}$\pm$0.019 & 0.477$\pm$0.092 & \textbf{0.345}$\pm$0.040 & 0.158$\pm$0.090 \\
		xalan-2.4$\Rightarrow$log4j-1.0 & 0.717$\pm$0.040 & \textbf{0.752}$\pm$0.026 & \textbf{0.477}$\pm$0.044 & 0.458$\pm$0.041 \\
		xerces-1.2$\Rightarrow$log4j-1.0 & \textbf{0.391}$\pm$0.125 & 0.202$\pm$0.077 & \textbf{0.098}$\pm$0.091 & -0.153$\pm$0.058 \\
		camel-1.0$\Rightarrow$ant-1.7 & 0.577$\pm$0.085 & \textbf{0.644}$\pm$0.069 & \textbf{0.338}$\pm$0.024 & 0.336$\pm$0.033 \\
		ivy-1.4$\Rightarrow$ant-1.7 & 0.459$\pm$0.114 & \textbf{0.531}$\pm$0.122 & 0.166$\pm$0.078 & \textbf{0.167}$\pm$0.106 \\
		tomcat$\Rightarrow$ant-1.7 & \textbf{0.575}$\pm$0.128 & 0.363$\pm$0.195 & \textbf{0.286}$\pm$0.074 & 0.189$\pm$0.098 \\
		prop-6$\Rightarrow$ant-1.7 & 0.223$\pm$0.138 & \textbf{0.263}$\pm$0.097 & \textbf{0.095}$\pm$0.045 & 0.062$\pm$0.060 \\
		poi-2.0$\Rightarrow$ant-1.7 & \textbf{0.679}$\pm$0.015 & 0.613$\pm$0.072 & \textbf{0.312}$\pm$0.022 & 0.265$\pm$0.049 \\
		xalan-2.4$\Rightarrow$ant-1.7 & 0.712$\pm$0.009 & \textbf{0.718}$\pm$0.012 & 0.380$\pm$0.018 & \textbf{0.383}$\pm$0.014 \\
		xerces-1.2$\Rightarrow$ant-1.7 & \textbf{0.440}$\pm$0.069 & 0.234$\pm$0.084 & \textbf{0.068}$\pm$0.057 & -0.082$\pm$0.059 \\
		camel-1.0$\Rightarrow$xerces-1.2 & 0.398$\pm$0.032 & \textbf{0.428}$\pm$0.028 & \textbf{0.127}$\pm$0.043 & 0.070$\pm$0.087 \\
		ivy-1.4$\Rightarrow$xerces-1.2 & 0.443$\pm$0.096 & \textbf{0.548}$\pm$0.100 & 0.130$\pm$0.051 & \textbf{0.178}$\pm$0.065 \\
		tomcat$\Rightarrow$xerces-1.2 & \textbf{0.457}$\pm$0.042 & 0.362$\pm$0.138 & \textbf{-0.023}$\pm$0.079 & -0.109$\pm$0.072 \\
		prop-6$\Rightarrow$xerces-1.2 & 0.159$\pm$0.112 & \textbf{0.161}$\pm$0.097 & \textbf{0.052}$\pm$0.048 & 0.040$\pm$0.054 \\
		poi-2.0$\Rightarrow$xerces-1.2 & 0.460$\pm$0.022 & \textbf{0.464}$\pm$0.025 & \textbf{0.074}$\pm$0.022 & 0.026$\pm$0.040 \\
		xalan-2.4$\Rightarrow$xerces-1.2 & 0.449$\pm$0.021 & \textbf{0.468}$\pm$0.012 & \textbf{0.122}$\pm$0.026 & 0.078$\pm$0.037 \\
		camel-1.0$\Rightarrow$velocity-1.6 & 0.286$\pm$0.060 & \textbf{0.400}$\pm$0.090 & 0.180$\pm$0.045 & \textbf{0.188}$\pm$0.039 \\
		ivy-1.4$\Rightarrow$velocity-1.6 & 0.303$\pm$0.116 & \textbf{0.406}$\pm$0.120 & 0.031$\pm$0.054 & \textbf{0.061}$\pm$0.093 \\
		tomcat$\Rightarrow$velocity-1.6 & \textbf{0.555}$\pm$0.130 & 0.374$\pm$0.204 & \textbf{0.247}$\pm$0.063 & 0.208$\pm$0.082 \\
		prop-6$\Rightarrow$velocity-1.6 & \textbf{0.242}$\pm$0.137 & 0.135$\pm$0.102 & 0.077$\pm$0.062 & \textbf{0.083}$\pm$0.068 \\
		poi-2.0$\Rightarrow$velocity-1.6 & 0.515$\pm$0.033 & \textbf{0.553}$\pm$0.092 & 0.178$\pm$0.028 & \textbf{0.207}$\pm$0.069 \\
		xalan-2.4$\Rightarrow$velocity-1.6 & 0.489$\pm$0.072 & \textbf{0.587}$\pm$0.040 & 0.195$\pm$0.052 & \textbf{0.256}$\pm$0.031 \\
		xerces-1.2$\Rightarrow$velocity-1.6 & \textbf{0.378}$\pm$0.098 & 0.148$\pm$0.087 & \textbf{0.112}$\pm$0.073 & -0.004$\pm$0.059 \\
		\textbf{Win/Tie/Lose} & ---  &  14/6/14 &   &  17/14/3 \\
		\textbf{Average} & 0.474$\pm$0.156 & 0.427$\pm$0.195 & 0.208$\pm$0.129 & 0.161$\pm$0.149 \\

		\hline
	\end{tabular}
\end{table}

%

Tables \ref{tb:p-deltaGM_MCC02} shows the results of P-value and Cliff's delta when comparing FWTNB and TNB on each dataset in terms of G-Measure and MCC, respectively. 


\begin{table}[htb] %
	\setlength{\abovecaptionskip}{0pt}
	\setlength{\belowcaptionskip}{10pt}
	\caption{\textbf{P-value} and \textbf{Cliff's delta} When Comparing our Proposed FETNB with TNB}
	\label{tb:p-deltaGM_MCC02}
	\centering
	\footnotesize
	\begin{tabular}{lllll}
		\hline
		\multirow{3}{*}{Source$\Rightarrow$Target} & \multicolumn{4}{c}{vs TNB} \\ 
		& \multicolumn{2}{c}{G-Measure} & \multicolumn{2}{c}{MCC} \\
		& P-Value & $\delta$ & P-Value & $\delta$ \\ \hline
		camel-1.0$\Rightarrow$jedit-4.0 & 0.095 & -0.252 & 0.165 & 0.210 \\
		ivy-1.4$\Rightarrow$jedit-4.0 & 0.025 & -0.338 & 0.149 & 0.218 \\
		tomcat$\Rightarrow$jedit-4.0 & 8.12e-4 & 0.504 & 1.41e-4 & 0.573 \\
		prop-6$\Rightarrow$jedit-4.0 & 8.65e-8 & 0.806 & 3.97e-9 & 0.886 \\
		poi-2.0$\Rightarrow$jedit-4.0 & 9.66e-9 & 0.863 & 9.49e-5 & 0.588 \\
		xalan-2.4$\Rightarrow$jedit-4.0 & 0.102 & 0.247 & 0.002 & 0.462 \\
		xerces-1.2$\Rightarrow$jedit-4.0 & 4.03e-11 & 0.993 & 8.07e-11 & 0.978 \\
		camel-1.0$\Rightarrow$log4j-1.0 & 0.001 & -0.493 & 0.695 & -0.06 \\
		ivy-1.4$\Rightarrow$log4j-1.0 & 0.007 & -0.408 & 0.779 & 0.043 \\
		tomcat$\Rightarrow$log4j-1.0 & 7.05e-8 & 0.811 & 1.15e-7 & 0.798 \\
		prop-6$\Rightarrow$log4j-1.0 & 0.359 & 0.139 & 0.147 & -0.219 \\
		poi-2.0$\Rightarrow$log4j-1.0 & 3.47e-8 & 0.830 & 5.19e-9 & 0.879 \\
		xalan-2.4$\Rightarrow$log4j-1.0 & 3.36e-4 & -0.54 & 0.162 & 0.211 \\
		xerces-1.2$\Rightarrow$log4j-1.0 & 9.48e-8 & 0.802 & 3.16e-11 & 0.998 \\
		camel-1.0$\Rightarrow$ant-1.7 & 0.001 & -0.496 & 0.6 & -0.08 \\
		ivy-1.4$\Rightarrow$ant-1.7 & 0.01 & -0.389 & 0.923 & -0.016 \\
		tomcat$\Rightarrow$ant-1.7 & 3.16e-5 & 0.627 & 1.41e-4 & 0.573 \\
		prop-6$\Rightarrow$ant-1.7 & 0.201 & -0.193 & 0.059 & 0.284 \\
		poi-2.0$\Rightarrow$ant-1.7 & 1.93e-8 & 0.846 & 3.95e-5 & 0.619 \\
		xalan-2.4$\Rightarrow$ant-1.7 & 0.013 & -0.373 & 0.387 & -0.131 \\
		xerces-1.2$\Rightarrow$ant-1.7 & 2.37e-10 & 0.953 & 8.89e-10 & 0.922 \\
		camel-1.0$\Rightarrow$xerces-1.2 & 1.73e-4 & -0.566 & 0.019 & 0.354 \\
		ivy-1.4$\Rightarrow$xerces-1.2 & 5.26e-5 & -0.609 & 0.01 & -0.389 \\
		tomcat$\Rightarrow$xerces-1.2 & 1.21e-4 & 0.579 & 1.54e-4 & 0.570 \\
		prop-6$\Rightarrow$xerces-1.2 & 0.865 & -0.027 & 0.34 & 0.144 \\
		poi-2.0$\Rightarrow$xerces-1.2 & 0.352 & -0.141 & 4.25e-6 & 0.692 \\
		xalan-2.4$\Rightarrow$xerces-1.2 & 2.12e-4 & -0.558 & 5.82e-6 & 0.682 \\
		camel-1.0$\Rightarrow$velocity-1.6 & 9.17e-6 & -0.668 & 0.261 & -0.17 \\
		ivy-1.4$\Rightarrow$velocity-1.6 & 0.001 & -0.483 & 0.287 & -0.161 \\
		tomcat$\Rightarrow$velocity-1.6 & 3.56e-4 & 0.538 & 0.045 & 0.302 \\
		prop-6$\Rightarrow$velocity-1.6 & 0.004 & 0.438 & 0.464 & -0.111 \\
		poi-2.0$\Rightarrow$velocity-1.6 & 0.003 & -0.443 & 0.019 & -0.352 \\
		xalan-2.4$\Rightarrow$velocity-1.6 & 1.94e-7 & -0.783 & 2.40e-6 & -0.71 \\
		xerces-1.2$\Rightarrow$velocity-1.6 & 3.06e-9 & 0.891 & 2.12e-7 & 0.780 \\
		
		\hline
	\end{tabular}	
\end{table}

\subsection{RQ3: Does TOMOFWTNB outperform other CPDP approaches?}


Tables \ref{tb:metricsGM03} and \ref{tb:metricsMCC03} respectively present G-Measure and MCC values in the fashion of 'mean$\pm$std' for our proposed TOMOFWTNB approach and baselines (i.e., seven previous CPDP approaches including Burak Filter \cite{Turhan2009On}, TNB \cite{Ma2012Transfer}, TCA+ \cite{Nam2013Transfer}, CCA+ \cite{Jing2015Heterogeneous}, HISNN \cite{Ryu2015A}, DBN-CP \cite{Wang2016Automatically}, and CTKCCA \cite{Li2018Cost}). The best value on each dataset is in bold. The average row means the average performance across all 34 groups of experiments. The row Win/Tie/Lose presents the results of comparing TOMOFWTNB with baselines based on the Wilcoxon rank-sum test. Win/Tie/Lose represents the number of datasets TOMOWFTNB wins/tie/loses, compared with the corresponding baseline.

From Tables \ref{tb:metricsGM03} and \ref{tb:metricsMCC03}, we notice that \textbf{TOMOFWTNB improves the average G-Measure value by 27.8\%$\sim$351.97\%, and the average MCC value by 71.5\%$\sim$1593.8\%, compared with the baselines}. 

More specifically, for G-Measure, TOMOFWTNB wins 23 datasets, ties 3 datasets, and loses 8 datasets, compared with Burak Filter; TOMOFWTNB wins 22 datasets, ties 4 datasets, and loses 8 datasets, compared with TNB; TOMOFWTNB wins 33 datasets and loses 1 dataset, compared with TCA+; TOMOFWTNB wins 33 datasets and ties 4 datasets, compared with CCA+, HISNN or DBN-CP; TOMOFWTNB wins 26 datasets, ties 4 datasets, and loses 4 datasets, compared with CTKCCA. With respect to MCC, TOMOFWTNB wins 26 datasets, ties 4 datasets, and loses 4 datasets, compared with Burak Filter; TOMOFWTNB wins 24 datasets, ties 6 datasets, and loses 4 datasets, compared with TNB; TOMOFWTNB wins 20 datasets, ties 1 dataset, and loses 3 datasets, compared with TCA+; TOMOFWTNB wins all 34 datasets compared with CCA+; TOMOFWTNB wins 29 datasets and ties 5 datasets, compared with TCA+; TOMOFWTNB wins 30 datasets, ties 3 datasets, and loses 1 dataset, compared with DBN-CP and CTKCCA.

The reason that the proposed TOMOFWTNB can achieve such good performance is that we take both class-imbalance problem and feature importance into consideration in CPDP scenario. We think the distribution differences between the source and target datasets should be taken into consideration when to address the class-imbalance problem in the CPDP scenario. Therefore, we propose a transfer-learning oriented minority over-sampling approach (TOMO), which can not only balance the imbalanced dataset but reduce the distribution difference between two datasets, to address the class-imbalance problem in CPDP scenario. Considering the fact that different features usually have different importance for predicting software defect-proneness, we propose a feature weighting transfer naive Bayes approach (FWTNB) based on TNB. Based on the above two work, a novel CPDP approach named TOMOFWTNB is proposed. We also notice that the proposed TOMOFWTNB achieves poor performance on some datasets compared with some baselines, e.g., camel-1.0$\Rightarrow$ant-1.7 TCA+. Possible reasons include: (1) Our TOMOFWTNB has two hyper-parameters $\lambda$ (see Step-11 in Algorithm \ref{alg:TOMO}) and $\sigma$ (see Eq.\ref{eq:fw_postProb}), which have an effect on the performance of TOMOFWTNB. However, in this paper, we just set $\lambda$ to 0.4, and $\sigma$ to 1, respectively. (2) The performance of cluster algorithm (see Step7 in Algorithm \ref{alg:TOMO}) used in TOMO will affect the performance of TOMOFWTNB, since no cluster algorithm always perform well on every dataset.


\begin{table}[htbp] %
	\setlength{\abovecaptionskip}{0pt}
	\setlength{\belowcaptionskip}{10pt}
	\renewcommand{\arraystretch}{1} 
	\caption{\textbf{G-Measure} of the proposed TOMOFWTNB and previous cross-project defect prediction approaches. The best value is in bold.}
	\label{tb:metricsGM03}
	\footnotesize
	\setlength{\leftskip}{-3.0cm}
	\begin{tabular}{lllllllll}
		\hline
		Source$\Rightarrow$Target & FWTNB & Burak Filter & TNB & TCA+ & CCA+ & HISNN & DBN-CP & CTKCCA \\ \hline
		camel-1.0$\Rightarrow$jedit-4.0 & 0.368$\pm$0.052 & 0.375$\pm$0.081 & \textbf{0.464}$\pm$0.147 & 0.443$\pm$0.073 & 0.077$\pm$0.054 & 0.144$\pm$0.086 & 0.010$\pm$0.019 & 0.425$\pm$0.006 \\
		ivy-1.4$\Rightarrow$jedit-4.0 & \textbf{0.662}$\pm$0.061 & 0.638$\pm$0.099 & 0.384$\pm$0.143 & 0.126$\pm$0.048 & 0.082$\pm$0.046 & 0.064$\pm$0.062 & 0.049$\pm$0.040 & 0.427$\pm$0.005 \\
		tomcat$\Rightarrow$jedit-4.0 & 0.699$\pm$0.009 & 0.033$\pm$0.032 & \textbf{0.720}$\pm$0.014 & 0.011$\pm$0.022 & 0.145$\pm$0.064 & 0.232$\pm$0.053 & 0.180$\pm$0.092 & 0.396$\pm$0.009 \\
		prop-6$\Rightarrow$jedit-4.0 & \textbf{0.685}$\pm$0.016 & 0.291$\pm$0.194 & 0.325$\pm$0.120 & 0.398$\pm$0.057 & 0.198$\pm$0.071 & 0.451$\pm$0.099 & 0.013$\pm$0.022 & 0.314$\pm$0.005 \\
		poi-2.0$\Rightarrow$jedit-4.0 & \textbf{0.685}$\pm$0.014 & 0.580$\pm$0.044 & 0.559$\pm$0.041 & 0.205$\pm$0.076 & 0.265$\pm$0.070 & 0.353$\pm$0.078 & 0.167$\pm$0.088 & 0.401$\pm$0.009 \\
		xalan-2.4$\Rightarrow$jedit-4.0 & 0.681$\pm$0.012 & 0.674$\pm$0.009 & \textbf{0.702}$\pm$0.014 & 0.164$\pm$0.038 & 0.296$\pm$0.074 & 0.404$\pm$0.049 & 0.122$\pm$0.040 & 0.426$\pm$0.009 \\
		xerces-1.2$\Rightarrow$jedit-4.0 & \textbf{0.632}$\pm$0.038 & 0.627$\pm$0.032 & 0.308$\pm$0.204 & 0.143$\pm$0.045 & 0.250$\pm$0.061 & 0.249$\pm$0.072 & 0.172$\pm$0.137 & 0.335$\pm$0.007 \\
		camel-1.0$\Rightarrow$log4j-1.0 & 0.556$\pm$0.088 & \textbf{0.598}$\pm$0.071 & 0.227$\pm$0.120 & 0.156$\pm$0.070 & 0.072$\pm$0.050 & 0.018$\pm$0.047 & 0.049$\pm$0.072 & 0.452$\pm$0.012 \\
		ivy-1.4$\Rightarrow$log4j-1.0 & \textbf{0.731}$\pm$0.037 & 0.533$\pm$0.083 & 0.367$\pm$0.062 & 0.013$\pm$0.028 & 0.120$\pm$0.074 & 0.101$\pm$0.116 & 0.172$\pm$0.163 & 0.585$\pm$0.007 \\
		tomcat$\Rightarrow$log4j-1.0 & \textbf{0.684}$\pm$0.055 & 0.191$\pm$0.241 & 0.643$\pm$0.032 & 0$\pm$0 & 0.201$\pm$0.078 & 0.002$\pm$0.010 & 0.447$\pm$0.174 & 0.554$\pm$0.012 \\
		prop-6$\Rightarrow$log4j-1.0 & \textbf{0.726}$\pm$0.046 & 0.458$\pm$0.218 & 0.515$\pm$0.054 & 0.052$\pm$0.045 & 0.192$\pm$0.096 & 0.153$\pm$0.060 & 0.081$\pm$0.133 & 0.483$\pm$0.011 \\
		poi-2.0$\Rightarrow$log4j-1.0 & \textbf{0.681}$\pm$0.042 & 0.649$\pm$0.041 & 0.457$\pm$0.091 & 0.019$\pm$0.031 & 0.205$\pm$0.085 & 0.217$\pm$0.067 & 0.439$\pm$0.147 & 0.509$\pm$0.014 \\
		xalan-2.4$\Rightarrow$log4j-1.0 & \textbf{0.746}$\pm$0.016 & 0.729$\pm$0.022 & 0.722$\pm$0.033 & 0.061$\pm$0.156 & 0.261$\pm$0.091 & 0.066$\pm$0.040 & 0.511$\pm$0.155 & 0.499$\pm$0.012 \\
		xerces-1.2$\Rightarrow$log4j-1.0 & 0.590$\pm$0.080 & 0.288$\pm$0.120 & 0.323$\pm$0.090 & 0.027$\pm$0.049 & 0.234$\pm$0.085 & 0.023$\pm$0.035 & 0.177$\pm$0.115 & \textbf{0.617}$\pm$0.010 \\
		camel-1.0$\Rightarrow$ant-1.7 & 0.387$\pm$0.096 & \textbf{0.502}$\pm$0.081 & 0.393$\pm$0.147 & 0.355$\pm$0.057 & 0.068$\pm$0.039 & 0.186$\pm$0.055 & 0.003$\pm$0.009 & 0.203$\pm$0.002 \\
		ivy-1.4$\Rightarrow$ant-1.7 & \textbf{0.697}$\pm$0.027 & 0.671$\pm$0.029 & 0.341$\pm$0.179 & 0.034$\pm$0.037 & 0.097$\pm$0.047 & 0.108$\pm$0.037 & 0.093$\pm$0.111 & 0.181$\pm$0.004 \\
		tomcat$\Rightarrow$ant-1.7 & 0.694$\pm$0.021 & 0.143$\pm$0.227 & \textbf{0.722}$\pm$0.016 & 0.058$\pm$0.013 & 0.173$\pm$0.055 & 0.195$\pm$0.029 & 0.356$\pm$0.164 & 0.265$\pm$0.006 \\
		prop-6$\Rightarrow$ant-1.7 & \textbf{0.686}$\pm$0.022 & 0.348$\pm$0.079 & 0.290$\pm$0.127 & 0.140$\pm$0.079 & 0.210$\pm$0.055 & 0.205$\pm$0.084 & 0.046$\pm$0.062 & 0.162$\pm$0.005 \\
		poi-2.0$\Rightarrow$ant-1.7 & \textbf{0.678}$\pm$0.013 & 0.613$\pm$0.029 & 0.511$\pm$0.034 & 0.145$\pm$0.030 & 0.240$\pm$0.049 & 0.365$\pm$0.074 & 0.263$\pm$0.158 & 0.210$\pm$0.004 \\
		xalan-2.4$\Rightarrow$ant-1.7 & 0.708$\pm$0.007 & 0.652$\pm$0.015 & \textbf{0.721}$\pm$0.009 & 0.095$\pm$0.015 & 0.283$\pm$0.053 & 0.304$\pm$0.028 & 0.174$\pm$0.084 & 0.218$\pm$0.005 \\
		xerces-1.2$\Rightarrow$ant-1.7 & \textbf{0.590}$\pm$0.052 & 0.192$\pm$0.083 & 0.266$\pm$0.093 & 0.101$\pm$0.023 & 0.257$\pm$0.047 & 0.054$\pm$0.034 & 0.141$\pm$0.109 & 0.176$\pm$0.005 \\
		camel-1.0$\Rightarrow$xerces-1.2 & 0.286$\pm$0.087 & 0.294$\pm$0.069 & \textbf{0.430}$\pm$0.140 & 0.248$\pm$0.061 & 0.062$\pm$0.062 & 0.128$\pm$0.024 & 0.019$\pm$0.035 & 0.252$\pm$0.019 \\
		ivy-1.4$\Rightarrow$xerces-1.2 & 0.443$\pm$0.028 & \textbf{0.498}$\pm$0.028 & 0.341$\pm$0.171 & 0.068$\pm$0.063 & 0.095$\pm$0.065 & 0.163$\pm$0.058 & 0.096$\pm$0.059 & 0.415$\pm$0.023 \\
		tomcat$\Rightarrow$xerces-1.2 & 0.450$\pm$0.014 & 0.089$\pm$0.015 & \textbf{0.496}$\pm$0.047 & 0.074$\pm$0.034 & 0.142$\pm$0.078 & 0.108$\pm$0.033 & 0.245$\pm$0.085 & 0.411$\pm$0.021 \\
		prop-6$\Rightarrow$xerces-1.2 & 0.379$\pm$0.037 & 0.288$\pm$0.145 & 0.253$\pm$0.133 & 0.130$\pm$0.086 & 0.185$\pm$0.109 & 0.185$\pm$0.032 & 0.044$\pm$0.044 & \textbf{0.408}$\pm$0.024 \\
		poi-2.0$\Rightarrow$xerces-1.2 & 0.493$\pm$0.022 & 0.528$\pm$0.013 & \textbf{0.556}$\pm$0.022 & 0.195$\pm$0.071 & 0.254$\pm$0.122 & 0.230$\pm$0.028 & 0.218$\pm$0.077 & 0.387$\pm$0.028 \\
		xalan-2.4$\Rightarrow$xerces-1.2 & 0.470$\pm$0.018 & \textbf{0.502}$\pm$0.010 & 0.460$\pm$0.006 & 0.114$\pm$0.034 & 0.279$\pm$0.100 & 0.213$\pm$0.037 & 0.137$\pm$0.051 & 0.427$\pm$0.017 \\
		camel-1.0$\Rightarrow$velocity-1.6 & 0.331$\pm$0.059 & 0.516$\pm$0.045 & 0.313$\pm$0.106 & 0.233$\pm$0.065 & 0.072$\pm$0.038 & 0.199$\pm$0.030 & 0.085$\pm$0.038 & \textbf{0.543}$\pm$0.008 \\
		ivy-1.4$\Rightarrow$velocity-1.6 & 0.509$\pm$0.049 & 0.418$\pm$0.029 & 0.279$\pm$0.196 & 0.109$\pm$0.041 & 0.082$\pm$0.038 & 0.146$\pm$0.036 & 0.072$\pm$0.048 & \textbf{0.527}$\pm$0.010 \\
		tomcat$\Rightarrow$velocity-1.6 & 0.498$\pm$0.062 & 0.228$\pm$0.251 & \textbf{0.585}$\pm$0.029 & 0.026$\pm$0.026 & 0.165$\pm$0.068 & 0.037$\pm$0.014 & 0.140$\pm$0.054 & 0.504$\pm$0.011 \\
		prop-6$\Rightarrow$velocity-1.6 & 0.434$\pm$0.051 & 0.267$\pm$0.091 & 0.279$\pm$0.145 & 0.079$\pm$0.048 & 0.185$\pm$0.065 & 0.113$\pm$0.027 & 0.033$\pm$0.048 & \textbf{0.541}$\pm$0.011 \\
		poi-2.0$\Rightarrow$velocity-1.6 & 0.561$\pm$0.039 & \textbf{0.618}$\pm$0.012 & 0.484$\pm$0.030 & 0.092$\pm$0.059 & 0.243$\pm$0.071 & 0.115$\pm$0.030 & 0.261$\pm$0.124 & 0.531$\pm$0.010 \\
		xalan-2.4$\Rightarrow$velocity-1.6 & 0.553$\pm$0.049 & \textbf{0.601}$\pm$0.016 & 0.567$\pm$0.030 & 0.107$\pm$0.130 & 0.251$\pm$0.070 & 0.120$\pm$0.010 & 0.096$\pm$0.028 & 0.479$\pm$0.007 \\
		xerces-1.2$\Rightarrow$velocity-1.6 & 0.534$\pm$0.054 & 0.327$\pm$0.049 & 0.272$\pm$0.166 & 0.105$\pm$0.082 & 0.222$\pm$0.059 & 0.079$\pm$0.034 & 0.076$\pm$0.054 & \textbf{0.556}$\pm$0.007 \\
		\textbf{Win/Tie/Lose} & --- &  23/3/8 &  22/4/8 &  33/0/1 &  34/0/0 &  34/0/0 &  34/0/0 &  26/4/4 \\
		\textbf{Average} & 0.574$\pm$0.132 & 0.440$\pm$0.192 & 0.449$\pm$0.157 & 0.127$\pm$0.107 & 0.181$\pm$0.075 & 0.169$\pm$0.111 & 0.153$\pm$0.129 & 0.406$\pm$0.131 \\

		\hline
	\end{tabular}
\end{table}

\begin{table}[htbp] %
	\setlength{\abovecaptionskip}{0pt}
	\setlength{\belowcaptionskip}{10pt}
	\renewcommand{\arraystretch}{1.0} 
	\caption{ \textbf{MCC} of the proposed TOMOFWTNB and Previous Cross-project Defect Prediction Models. The best value is in bold.}
	\label{tb:metricsMCC03}
	\footnotesize
	\setlength{\leftskip}{-3.1cm}
	\begin{tabular}{lllllllll}
		\hline
		Source$\Rightarrow$Target & FWTNB & Burak Filter & TNB & TCA+ & CCA+ & HISNN & DBN-CP & CTKCCA \\ \hline
		camel-1.0$\Rightarrow$jedit-4.0 & 0.170$\pm$0.040 & 0.047$\pm$0.072 & 0.170$\pm$0.059 & \textbf{0.284}$\pm$0.057 & 0.035$\pm$0.078 & 0.075$\pm$0.071 & 0.001$\pm$0.038 & 0.078$\pm$0.007 \\
		ivy-1.4$\Rightarrow$jedit-4.0 & \textbf{0.339}$\pm$0.068 & 0.324$\pm$0.100 & 0.195$\pm$0.060 & 0.113$\pm$0.044 & 0.026$\pm$0.070 & -0.043$\pm$0.060 & 0.075$\pm$0.081 & 0.075$\pm$0.009 \\
		tomcat$\Rightarrow$jedit-4.0 & \textbf{0.428}$\pm$0.020 & 0.049$\pm$0.018 & 0.391$\pm$0.024 & 0.012$\pm$0.041 & -0.001$\pm$0.074 & 0.248$\pm$0.044 & 0.211$\pm$0.085 & 0.006$\pm$0.018 \\
		prop-6$\Rightarrow$jedit-4.0 & \textbf{0.392}$\pm$0.031 & 0.103$\pm$0.095 & 0.175$\pm$0.058 & 0.251$\pm$0.055 & -0.002$\pm$0.076 & 0.350$\pm$0.060 & 0.001$\pm$0.046 & -0.054$\pm$0.006 \\
		poi-2.0$\Rightarrow$jedit-4.0 & \textbf{0.332}$\pm$0.025 & 0.190$\pm$0.056 & 0.293$\pm$0.016 & 0.163$\pm$0.084 & 0.035$\pm$0.072 & 0.206$\pm$0.067 & 0.077$\pm$0.066 & 0.039$\pm$0.013 \\
		xalan-2.4$\Rightarrow$jedit-4.0 & \textbf{0.374}$\pm$0.019 & 0.331$\pm$0.010 & 0.363$\pm$0.027 & 0.153$\pm$0.032 & 0.030$\pm$0.079 & 0.252$\pm$0.031 & 0.123$\pm$0.036 & 0.077$\pm$0.010 \\
		xerces-1.2$\Rightarrow$jedit-4.0 & 0.247$\pm$0.062 & \textbf{0.314}$\pm$0.020 & 0.087$\pm$0.069 & 0.107$\pm$0.083 & 0.005$\pm$0.059 & 0.163$\pm$0.060 & 0.088$\pm$0.096 & -0.043$\pm$0.016 \\
		camel-1.0$\Rightarrow$log4j-1.0 & \textbf{0.469}$\pm$0.071 & 0.255$\pm$0.066 & -0.111$\pm$0.098 & 0.208$\pm$0.084 & 0.016$\pm$0.099 & -0.001$\pm$0.070 & 0.069$\pm$0.090 & -0.082$\pm$0.020 \\
		ivy-1.4$\Rightarrow$log4j-1.0 & \textbf{0.520}$\pm$0.043 & 0.322$\pm$0.087 & -0.127$\pm$0.057 & 0.030$\pm$0.060 & 0.049$\pm$0.102 & -0.026$\pm$0.094 & 0.156$\pm$0.168 & 0.155$\pm$0.014 \\
		tomcat$\Rightarrow$log4j-1.0 & \textbf{0.434}$\pm$0.047 & 0.145$\pm$0.164 & 0.282$\pm$0.059 & 0$\pm$0 & 0.046$\pm$0.073 & 0.005$\pm$0.027 & 0.363$\pm$0.136 & 0.102$\pm$0.022 \\
		prop-6$\Rightarrow$log4j-1.0 & \textbf{0.458}$\pm$0.051 & 0.128$\pm$0.209 & 0.121$\pm$0.050 & 0.082$\pm$0.076 & -0.003$\pm$0.102 & 0.183$\pm$0.059 & 0.065$\pm$0.113 & -0.029$\pm$0.019 \\
		poi-2.0$\Rightarrow$log4j-1.0 & 0.368$\pm$0.070 & \textbf{0.378}$\pm$0.034 & -0.005$\pm$0.071 & 0.042$\pm$0.073 & -0.022$\pm$0.093 & 0.291$\pm$0.043 & 0.155$\pm$0.125 & 0.016$\pm$0.024 \\
		xalan-2.4$\Rightarrow$log4j-1.0 & \textbf{0.473}$\pm$0.048 & 0.450$\pm$0.028 & 0.425$\pm$0.039 & 0.054$\pm$0.112 & 0.010$\pm$0.078 & 0.123$\pm$0.068 & 0.243$\pm$0.116 & 0$\pm$0.022 \\
		xerces-1.2$\Rightarrow$log4j-1.0 & \textbf{0.245}$\pm$0.108 & 0.100$\pm$0.082 & -0.165$\pm$0.056 & -0.012$\pm$0.136 & -0.008$\pm$0.080 & -0.05$\pm$0.043 & 0.159$\pm$0.093 & 0.218$\pm$0.018 \\
		camel-1.0$\Rightarrow$ant-1.7 & 0.144$\pm$0.053 & 0.088$\pm$0.048 & 0.067$\pm$0.102 & \textbf{0.226}$\pm$0.059 & 0.039$\pm$0.051 & 0.127$\pm$0.050 & -0.003$\pm$0.038 & 0.003$\pm$0.004 \\
		ivy-1.4$\Rightarrow$ant-1.7 & \textbf{0.360}$\pm$0.037 & 0.347$\pm$0.036 & 0.025$\pm$0.118 & 0.043$\pm$0.042 & 0.032$\pm$0.052 & 0.036$\pm$0.037 & 0.110$\pm$0.118 & -0.045$\pm$0.006 \\
		tomcat$\Rightarrow$ant-1.7 & 0.378$\pm$0.028 & 0.061$\pm$0.126 & \textbf{0.388}$\pm$0.028 & 0.151$\pm$0.019 & 0.036$\pm$0.041 & 0.196$\pm$0.025 & 0.295$\pm$0.079 & 0.066$\pm$0.021 \\
		prop-6$\Rightarrow$ant-1.7 & \textbf{0.328}$\pm$0.037 & 0.006$\pm$0.085 & 0.081$\pm$0.069 & 0.105$\pm$0.062 & 0.019$\pm$0.053 & 0.215$\pm$0.060 & 0.078$\pm$0.079 & -0.053$\pm$0.009 \\
		poi-2.0$\Rightarrow$ant-1.7 & 0.310$\pm$0.021 & 0.211$\pm$0.044 & 0.067$\pm$0.041 & 0.197$\pm$0.028 & 0.004$\pm$0.048 & \textbf{0.316}$\pm$0.037 & 0.203$\pm$0.071 & 0.026$\pm$0.005 \\
		xalan-2.4$\Rightarrow$ant-1.7 & 0.361$\pm$0.016 & 0.328$\pm$0.012 & \textbf{0.383}$\pm$0.016 & 0.174$\pm$0.018 & 0.021$\pm$0.060 & 0.272$\pm$0.024 & 0.198$\pm$0.064 & 0.024$\pm$0.007 \\
		xerces-1.2$\Rightarrow$ant-1.7 & \textbf{0.184}$\pm$0.090 & 0.051$\pm$0.039 & -0.098$\pm$0.021 & 0.156$\pm$0.025 & 0.025$\pm$0.057 & 0.024$\pm$0.038 & 0.073$\pm$0.084 & -0.016$\pm$0.006 \\
		camel-1.0$\Rightarrow$xerces-1.2 & 0.070$\pm$0.045 & -0.051$\pm$0.043 & \textbf{0.103}$\pm$0.092 & -0.047$\pm$0.040 & -0.002$\pm$0.059 & 0.007$\pm$0.041 & -0.025$\pm$0.043 & -0.074$\pm$0.012 \\
		ivy-1.4$\Rightarrow$xerces-1.2 & 0.092$\pm$0.025 & \textbf{0.186}$\pm$0.024 & 0.085$\pm$0.091 & -0.04$\pm$0.051 & 0.019$\pm$0.060 & 0$\pm$0.031 & 0.090$\pm$0.060 & 0.052$\pm$0.021 \\
		tomcat$\Rightarrow$xerces-1.2 & 0.130$\pm$0.015 & 0.054$\pm$0.021 & 0.084$\pm$0.025 & 0.049$\pm$0.037 & 0.011$\pm$0.063 & 0.085$\pm$0.040 & \textbf{0.150}$\pm$0.050 & 0.047$\pm$0.017 \\
		prop-6$\Rightarrow$xerces-1.2 & 0.062$\pm$0.024 & -0.154$\pm$0.168 & \textbf{0.083}$\pm$0.041 & 0.001$\pm$0.047 & 0.005$\pm$0.080 & 0.044$\pm$0.013 & 0.043$\pm$0.060 & 0.042$\pm$0.019 \\
		poi-2.0$\Rightarrow$xerces-1.2 & 0.141$\pm$0.022 & 0.106$\pm$0.015 & \textbf{0.158}$\pm$0.020 & -0.001$\pm$0.054 & 0.010$\pm$0.077 & 0.039$\pm$0.024 & 0.050$\pm$0.072 & 0.019$\pm$0.025 \\
		xalan-2.4$\Rightarrow$xerces-1.2 & 0.108$\pm$0.018 & 0.058$\pm$0.010 & 0.088$\pm$0.023 & 0.058$\pm$0.031 & 0.014$\pm$0.084 & \textbf{0.109}$\pm$0.029 & 0.045$\pm$0.038 & 0.065$\pm$0.016 \\
		camel-1.0$\Rightarrow$velocity-1.6 & 0.174$\pm$0.043 & \textbf{0.200}$\pm$0.035 & -0.091$\pm$0.123 & 0.133$\pm$0.072 & 0.033$\pm$0.059 & 0.194$\pm$0.041 & 0.106$\pm$0.046 & 0.160$\pm$0.012 \\
		ivy-1.4$\Rightarrow$velocity-1.6 & \textbf{0.227}$\pm$0.041 & 0.165$\pm$0.034 & -0.035$\pm$0.092 & 0.107$\pm$0.064 & 0.006$\pm$0.084 & 0.188$\pm$0.041 & 0.106$\pm$0.056 & 0.117$\pm$0.016 \\
		tomcat$\Rightarrow$velocity-1.6 & 0.230$\pm$0.046 & 0.044$\pm$0.165 & \textbf{0.234}$\pm$0.059 & 0.068$\pm$0.066 & 0.021$\pm$0.082 & 0.065$\pm$0.049 & 0.163$\pm$0.042 & 0.090$\pm$0.019 \\
		prop-6$\Rightarrow$velocity-1.6 & 0.143$\pm$0.050 & -0.024$\pm$0.099 & -0.014$\pm$0.065 & 0.079$\pm$0.044 & 0.018$\pm$0.074 & 0.074$\pm$0.039 & 0.045$\pm$0.063 & \textbf{0.145}$\pm$0.022 \\
		poi-2.0$\Rightarrow$velocity-1.6 & 0.199$\pm$0.057 & \textbf{0.240}$\pm$0.015 & -0.006$\pm$0.068 & 0.113$\pm$0.037 & -0.013$\pm$0.079 & 0.045$\pm$0.029 & 0.196$\pm$0.068 & 0.133$\pm$0.017 \\
		xalan-2.4$\Rightarrow$velocity-1.6 & 0.217$\pm$0.031 & 0.202$\pm$0.026 & \textbf{0.270}$\pm$0.023 & 0.063$\pm$0.067 & 0.011$\pm$0.066 & 0.156$\pm$0.028 & 0.131$\pm$0.051 & 0.043$\pm$0.013 \\
		xerces-1.2$\Rightarrow$velocity-1.6 & 0.107$\pm$0.084 & 0.127$\pm$0.013 & -0.043$\pm$0.047 & 0.157$\pm$0.035 & 0.014$\pm$0.078 & 0.075$\pm$0.055 & 0.044$\pm$0.077 & \textbf{0.168}$\pm$0.014 \\
		\textbf{Win/Tie/Lose} & ---  &  26/4/4 &  24/6/4 &  30/1/3 &  34/0/0 &  29/5/0 &  30/3/1 &  30/3/1 \\
		\textbf{Average} & 0.271$\pm$0.133 & 0.158$\pm$0.137 & 0.115$\pm$0.163 & 0.096$\pm$0.083 & 0.016$\pm$0.017 & 0.119$\pm$0.109 & 0.114$\pm$0.086 & 0.046$\pm$0.075 \\

		\hline
	\end{tabular}
\end{table}

%
%

%

Tables \ref{tb:p-delta_GM03} and \ref{tb:p-delta_MCC03} show the results of P-value and Cliff's delta when comparing our proposed TOMOFWTNB and the baselines on each dataset in terms of G-Measure and MCC, respectively. 


\begin{table}[htb] %
	\setlength{\abovecaptionskip}{0pt}
	\setlength{\belowcaptionskip}{10pt}
	\renewcommand{\arraystretch}{1.0}
	\caption{ \textbf{P-value} and \textbf{Cliff's delta} When Comparing our Proposed TOFWTNB with the Baselines in terms of \textbf{G-Measure}}
	\label{tb:p-delta_GM03}
	\centering
	\footnotesize
	\setlength{\leftskip}{-3.9cm}
	\begin{tabular}{lllllllllllllll}
		\hline
		 \multirow{2}{*}{Source$\Rightarrow$Target} & \multicolumn{2}{c}{ vs Burak Filter} & \multicolumn{2}{c}{ vs TNB} & \multicolumn{2}{c}{ vs TCA+} & \multicolumn{2}{c}{ vs CCA+} & \multicolumn{2}{c}{ vs HISNN} & \multicolumn{2}{c}{ vs DBN-CP} & \multicolumn{2}{c}{vs CTKCCA} \\
		& P-Value & $\delta$ & P-Value & $\delta$ & P-Value & $\delta$ & P-Value & $\delta$ & P-Value & $\delta$ & P-Value & $\delta$ & P-Value & $\delta$ \\ \hline
		camel-1.0$\Rightarrow$jedit-4.0 & 0.842 & 0.031 & 0.395 & -0.129 & 1.75e-5 & -0.647 & 3.01e-11 & 1 & 2.37e-10 & 0.953 & 9.39e-12 & 1 & 1.28e-7 & -0.791 \\
		ivy-1.4$\Rightarrow$jedit-4.0 & 0.037 & 0.314 & 2.12e-10 & 0.947 & 2.75e-11 & 1 & 3.00e-11 & 1 & 2.89e-11 & 1 & 2.92e-11 & 1 & 4.81e-10 & 0.933 \\
		tomcat$\Rightarrow$jedit-4.0 & 4.10e-12 & 1 & 3.79e-7 & -0.764 & 9.33e-12 & 1 & 3.01e-11 & 1 & 2.99e-11 & 1 & 2.99e-11 & 1 & 2.96e-11 & 1 \\
		prop-6$\Rightarrow$jedit-4.0 & 1.96e-11 & 0.984 & 9.51e-11 & 0.973 & 3.01e-11 & 1 & 3.02e-11 & 1 & 3.01e-11 & 1 & 1.26e-11 & 1 & 2.40e-11 & 1 \\
		poi-2.0$\Rightarrow$jedit-4.0 & 9.73e-10 & 0.920 & 2.99e-11 & 1 & 2.97e-11 & 1 & 3.01e-11 & 1 & 3.01e-11 & 1 & 3.00e-11 & 1 & 2.72e-11 & 1 \\
		xalan-2.4$\Rightarrow$jedit-4.0 & 0.013 & 0.376 & 5.59e-7 & -0.753 & 2.94e-11 & 1 & 3.01e-11 & 1 & 3.00e-11 & 1 & 2.98e-11 & 1 & 2.82e-11 & 1 \\
		xerces-1.2$\Rightarrow$jedit-4.0 & 0.34 & 0.144 & 7.89e-11 & 0.978 & 2.10e-11 & 1 & 3.02e-11 & 1 & 3.02e-11 & 1 & 3.01e-11 & 1 & 2.89e-11 & 1 \\
		camel-1.0$\Rightarrow$log4j-1.0 & 0.024 & -0.34 & 5.36e-11 & 0.984 & 2.84e-11 & 1 & 2.88e-11 & 1 & 5.17e-12 & 1 & 1.87e-11 & 1 & 6.62e-6 & 0.673 \\
		ivy-1.4$\Rightarrow$log4j-1.0 & 3.37e-11 & 0.997 & 1.59e-11 & 1 & 6.33e-12 & 1 & 2.97e-11 & 1 & 2.24e-11 & 1 & 2.77e-11 & 1 & 2.63e-11 & 1 \\
		tomcat$\Rightarrow$log4j-1.0 & 1.16e-7 & 0.787 & 4.58e-4 & 0.528 & 1.21e-12 & 1 & 3.00e-11 & 1 & 1.72e-12 & 1 & 4.99e-9 & 0.880 & 3.99e-10 & 0.936 \\
		prop-6$\Rightarrow$log4j-1.0 & 8.61e-8 & 0.803 & 2.89e-11 & 1 & 2.50e-11 & 1 & 3.01e-11 & 1 & 2.66e-11 & 1 & 2.07e-11 & 1 & 2.69e-11 & 1 \\
		poi-2.0$\Rightarrow$log4j-1.0 & 0.01 & 0.389 & 2.92e-11 & 1 & 1.04e-11 & 1 & 3.01e-11 & 1 & 2.75e-11 & 1 & 4.07e-11 & 0.993 & 1.42e-11 & 1 \\
		xalan-2.4$\Rightarrow$log4j-1.0 & 2.76e-4 & 0.547 & 0.007 & 0.406 & 1.09e-11 & 1 & 3.01e-11 & 1 & 2.32e-11 & 1 & 7.37e-11 & 0.980 & 2.47e-11 & 1 \\
		xerces-1.2$\Rightarrow$log4j-1.0 & 2.55e-9 & 0.896 & 3.31e-11 & 0.991 & 1.10e-11 & 1 & 3.32e-11 & 0.998 & 1.26e-11 & 1 & 3.00e-11 & 1 & 0.129 & -0.229 \\
		camel-1.0$\Rightarrow$ant-1.7 & 3.81e-7 & -0.764 & 0.853 & 0.029 & 0.026 & 0.336 & 3.34e-11 & 0.998 & 5.46e-9 & 0.878 & 5.22e-12 & 1 & 5.25e-10 & 0.931 \\
		ivy-1.4$\Rightarrow$ant-1.7 & 2.15e-6 & 0.713 & 3.07e-11 & 0.996 & 2.93e-11 & 1 & 3.02e-11 & 1 & 3.01e-11 & 1 & 2.97e-11 & 1 & 1.92e-11 & 1 \\
		tomcat$\Rightarrow$ant-1.7 & 5.03e-8 & 0.820 & 1.73e-6 & -0.72 & 2.28e-11 & 1 & 3.02e-11 & 1 & 3.00e-11 & 1 & 3.02e-11 & 1 & 2.95e-11 & 1 \\
		prop-6$\Rightarrow$ant-1.7 & 1.85e-11 & 0.969 & 3.99e-11 & 0.993 & 3.01e-11 & 1 & 3.02e-11 & 1 & 3.00e-11 & 1 & 2.94e-11 & 1 & 2.90e-11 & 1 \\
		poi-2.0$\Rightarrow$ant-1.7 & 2.15e-10 & 0.956 & 3.01e-11 & 1 & 2.71e-11 & 1 & 3.02e-11 & 1 & 3.02e-11 & 1 & 3.01e-11 & 1 & 2.38e-11 & 1 \\
		xalan-2.4$\Rightarrow$ant-1.7 & 3.34e-11 & 0.998 & 1.61e-6 & -0.722 & 2.68e-11 & 1 & 3.02e-11 & 1 & 3.01e-11 & 1 & 3.02e-11 & 1 & 2.74e-11 & 1 \\
		xerces-1.2$\Rightarrow$ant-1.7 & 3.01e-11 & 1 & 2.25e-11 & 1 & 2.79e-11 & 1 & 3.02e-11 & 1 & 3.00e-11 & 1 & 3.02e-11 & 1 & 2.38e-11 & 1 \\
		camel-1.0$\Rightarrow$xerces-1.2 & 0.819 & 0.036 & 0.004 & -0.433 & 0.013 & 0.376 & 6.51e-10 & 0.929 & 3.49e-9 & 0.889 & 1.67e-11 & 0.991 & 0.003 & 0.449 \\
		ivy-1.4$\Rightarrow$xerces-1.2 & 1.35e-8 & -0.854 & 0.183 & 0.200 & 2.50e-11 & 1 & 2.98e-11 & 1 & 3.00e-11 & 1 & 3.00e-11 & 1 & 1.59e-6 & 0.717 \\
		tomcat$\Rightarrow$xerces-1.2 & 2.72e-11 & 1 & 2.86e-6 & -0.704 & 1.08e-11 & 1 & 3.00e-11 & 1 & 2.97e-11 & 1 & 3.00e-11 & 1 & 4.93e-10 & 0.936 \\
		prop-6$\Rightarrow$xerces-1.2 & 9.95e-4 & 0.487 & 0.004 & 0.436 & 1.09e-10 & 0.971 & 2.03e-9 & 0.902 & 2.97e-11 & 1 & 2.80e-11 & 1 & 0.005 & -0.421 \\
		poi-2.0$\Rightarrow$xerces-1.2 & 6.48e-9 & -0.873 & 2.36e-10 & -0.953 & 5.48e-11 & 0.987 & 1.29e-9 & 0.913 & 3.01e-11 & 1 & 3.02e-11 & 1 & 5.01e-11 & 0.987 \\
		xalan-2.4$\Rightarrow$xerces-1.2 & 3.60e-8 & -0.829 & 0.009 & 0.392 & 2.96e-11 & 1 & 5.54e-10 & 0.933 & 2.99e-11 & 1 & 2.99e-11 & 1 & 4.73e-10 & 0.937 \\
		camel-1.0$\Rightarrow$velocity-1.6 & 3.32e-11 & -0.998 & 0.013 & 0.373 & 1.49e-6 & 0.724 & 3.00e-11 & 1 & 2.00e-9 & 0.902 & 3.29e-11 & 0.998 & 2.87e-11 & -1 \\
		ivy-1.4$\Rightarrow$velocity-1.6 & 2.00e-8 & 0.844 & 1.00e-4 & 0.578 & 2.92e-11 & 1 & 3.00e-11 & 1 & 2.92e-11 & 1 & 2.93e-11 & 1 & 0.052 & -0.291 \\
		tomcat$\Rightarrow$velocity-1.6 & 0.009 & 0.394 & 2.03e-9 & -0.902 & 1.78e-11 & 1 & 4.07e-11 & 0.993 & 2.52e-11 & 1 & 2.98e-11 & 1 & 0.48 & 0.107 \\
		prop-6$\Rightarrow$velocity-1.6 & 2.31e-8 & 0.839 & 0.002 & 0.478 & 2.90e-11 & 1 & 3.01e-11 & 1 & 2.95e-11 & 1 & 2.26e-11 & 1 & 4.75e-11 & -0.989 \\
		poi-2.0$\Rightarrow$velocity-1.6 & 2.65e-9 & -0.896 & 1.19e-8 & 0.858 & 2.84e-11 & 1 & 3.01e-11 & 1 & 2.96e-11 & 1 & 3.31e-11 & 0.998 & 3.59e-4 & 0.537 \\
		xalan-2.4$\Rightarrow$velocity-1.6 & 2.86e-7 & -0.772 & 0.501 & -0.102 & 1.06e-10 & 0.971 & 3.66e-11 & 0.996 & 2.30e-11 & 1 & 2.87e-11 & 1 & 9.47e-9 & 0.862 \\
		xerces-1.2$\Rightarrow$velocity-1.6 & 1.14e-10 & 0.969 & 5.83e-9 & 0.871 & 5.03e-11 & 0.984 & 3.01e-11 & 1 & 2.90e-11 & 1 & 2.95e-11 & 1 & 0.072 & -0.271 \\

		\hline
	\end{tabular}	
\end{table}

\begin{table}[htb] %
	\setlength{\abovecaptionskip}{0pt}
	\setlength{\belowcaptionskip}{10pt}
	\renewcommand{\arraystretch}{1.0}
	\caption{\textbf{P-value} and \textbf{Cliff's delta} When Comparing our Proposed TOMOFWTNB with the Baselines in terms of \textbf{MCC}}
	\label{tb:p-delta_MCC03}
	\centering
	\footnotesize
	\setlength{\leftskip}{-3.9cm}
	\begin{tabular}{lllllllllllllll}
		\hline
		\multirow{2}{*}{Source$\Rightarrow$Target} & \multicolumn{2}{c}{ vs Burak Filter} & \multicolumn{2}{c}{ vs TNB} & \multicolumn{2}{c}{ vs TCA+} & \multicolumn{2}{c}{ vs CCA+} & \multicolumn{2}{c}{ vs HISNN} & \multicolumn{2}{c}{ vs DBN-CP} & \multicolumn{2}{c}{vs CTKCCA} \\
		& P-Value & $\delta$ & P-Value & $\delta$ & P-Value & $\delta$ & P-Value & $\delta$ & P-Value & $\delta$ & P-Value & $\delta$ & P-Value & $\delta$ \\ \hline
		camel-1.0$\Rightarrow$jedit-4.0 & 1.06e-7 & 0.800 & 0.34 & 0.144 & 1.20e-8 & -0.858 & 4.99e-9 & 0.880 & 4.11e-7 & 0.762 & 3.15e-11 & 0.991 & 5.41e-11 & 0.982 \\
		ivy-1.4$\Rightarrow$jedit-4.0 & 0.69 & 0.061 & 7.93e-9 & 0.860 & 1.63e-10 & 0.960 & 4.05e-11 & 0.993 & 3.30e-11 & 0.998 & 1.19e-10 & 0.969 & 2.55e-11 & 1 \\
		tomcat$\Rightarrow$jedit-4.0 & 4.10e-12 & 1 & 2.37e-7 & 0.778 & 2.02e-11 & 1 & 3.01e-11 & 1 & 2.99e-11 & 1 & 6.62e-11 & 0.982 & 2.96e-11 & 1 \\
		prop-6$\Rightarrow$jedit-4.0 & 9.40e-12 & 1 & 3.53e-11 & 0.996 & 4.96e-11 & 0.989 & 3.02e-11 & 1 & 0.002 & 0.464 & 1.94e-11 & 1 & 2.40e-11 & 1 \\
		poi-2.0$\Rightarrow$jedit-4.0 & 5.06e-10 & 0.936 & 2.18e-8 & 0.842 & 3.42e-10 & 0.944 & 3.01e-11 & 1 & 1.41e-9 & 0.911 & 3.00e-11 & 1 & 2.72e-11 & 1 \\
		xalan-2.4$\Rightarrow$jedit-4.0 & 5.43e-11 & 0.987 & 0.077 & 0.267 & 2.94e-11 & 1 & 3.01e-11 & 1 & 3.00e-11 & 1 & 2.98e-11 & 1 & 2.82e-11 & 1 \\
		xerces-1.2$\Rightarrow$jedit-4.0 & 7.04e-7 & -0.747 & 5.41e-10 & 0.933 & 1.66e-9 & 0.900 & 3.34e-11 & 0.998 & 9.51e-6 & 0.667 & 4.68e-8 & 0.822 & 2.89e-11 & 1 \\
		camel-1.0$\Rightarrow$log4j-1.0 & 1.18e-10 & 0.969 & 2.67e-11 & 1 & 3.48e-11 & 0.996 & 2.94e-11 & 1 & 1.99e-11 & 1 & 2.02e-11 & 1 & 2.16e-11 & 1 \\
		ivy-1.4$\Rightarrow$log4j-1.0 & 4.55e-11 & 0.990 & 1.59e-11 & 1 & 6.33e-12 & 1 & 2.98e-11 & 1 & 2.95e-11 & 1 & 3.28e-11 & 0.998 & 2.63e-11 & 1 \\
		tomcat$\Rightarrow$log4j-1.0 & 9.86e-7 & 0.727 & 8.52e-11 & 0.977 & 1.21e-12 & 1 & 3.00e-11 & 1 & 1.72e-12 & 1 & 0.03 & 0.327 & 2.29e-11 & 1 \\
		prop-6$\Rightarrow$log4j-1.0 & 1.35e-10 & 0.963 & 2.89e-11 & 1 & 2.60e-11 & 1 & 3.01e-11 & 1 & 2.66e-11 & 1 & 6.07e-11 & 0.980 & 2.69e-11 & 1 \\
		poi-2.0$\Rightarrow$log4j-1.0 & 0.971 & -0.007 & 2.92e-11 & 1 & 1.20e-11 & 1 & 3.33e-11 & 0.998 & 6.91e-6 & 0.676 & 4.99e-9 & 0.880 & 1.42e-11 & 1 \\
		xalan-2.4$\Rightarrow$log4j-1.0 & 0.048 & 0.298 & 0.001 & 0.494 & 1.43e-11 & 1 & 3.01e-11 & 1 & 2.32e-11 & 1 & 5.06e-10 & 0.936 & 2.47e-11 & 1 \\
		xerces-1.2$\Rightarrow$log4j-1.0 & 2.69e-7 & 0.773 & 2.21e-11 & 1 & 1.02e-9 & 0.899 & 5.55e-10 & 0.933 & 3.10e-11 & 0.998 & 0.006 & 0.412 & 0.454 & 0.113 \\
		camel-1.0$\Rightarrow$ant-1.7 & 5.97e-5 & 0.604 & 0.002 & 0.460 & 6.52e-7 & -0.749 & 3.08e-8 & 0.833 & 0.176 & 0.204 & 1.04e-10 & 0.964 & 2.53e-11 & 1 \\
		ivy-1.4$\Rightarrow$ant-1.7 & 0.105 & 0.244 & 3.07e-11 & 0.996 & 2.95e-11 & 1 & 3.02e-11 & 1 & 3.01e-11 & 1 & 7.34e-11 & 0.980 & 1.92e-11 & 1 \\
		tomcat$\Rightarrow$ant-1.7 & 3.11e-10 & 0.947 & 0.284 & -0.162 & 2.28e-11 & 1 & 3.02e-11 & 1 & 3.00e-11 & 1 & 2.49e-6 & 0.709 & 2.95e-11 & 1 \\
		prop-6$\Rightarrow$ant-1.7 & 4.75e-11 & 0.949 & 3.61e-11 & 0.996 & 5.48e-11 & 0.987 & 3.02e-11 & 1 & 1.01e-8 & 0.862 & 4.01e-11 & 0.993 & 2.90e-11 & 1 \\
		poi-2.0$\Rightarrow$ant-1.7 & 4.61e-10 & 0.938 & 3.01e-11 & 1 & 2.71e-11 & 1 & 3.02e-11 & 1 & 0.42 & -0.122 & 7.75e-9 & 0.869 & 2.38e-11 & 1 \\
		xalan-2.4$\Rightarrow$ant-1.7 & 8.10e-10 & 0.924 & 7.22e-6 & -0.676 & 2.68e-11 & 1 & 3.02e-11 & 1 & 3.01e-11 & 1 & 1.96e-10 & 0.958 & 2.74e-11 & 1 \\
		xerces-1.2$\Rightarrow$ant-1.7 & 3.63e-8 & 0.829 & 2.75e-11 & 0.996 & 0.067 & 0.276 & 1.43e-8 & 0.853 & 4.97e-9 & 0.880 & 2.00e-5 & 0.642 & 4.53e-10 & 0.933 \\
		camel-1.0$\Rightarrow$xerces-1.2 & 5.57e-10 & 0.933 & 0.5 & -0.102 & 1.46e-10 & 0.964 & 2.28e-5 & 0.638 & 3.83e-6 & 0.696 & 1.16e-8 & 0.858 & 2.59e-11 & 1 \\
		ivy-1.4$\Rightarrow$xerces-1.2 & 5.14e-11 & -0.988 & 0.193 & 0.196 & 4.88e-11 & 0.989 & 3.24e-7 & 0.769 & 8.10e-11 & 0.978 & 0.455 & -0.113 & 8.03e-8 & 0.801 \\
		tomcat$\Rightarrow$xerces-1.2 & 4.06e-11 & 0.991 & 4.53e-9 & 0.882 & 1.09e-11 & 1 & 9.69e-10 & 0.920 & 2.14e-6 & 0.713 & 0.003 & -0.442 & 2.93e-11 & 1 \\
		prop-6$\Rightarrow$xerces-1.2 & 1.42e-4 & 0.562 & 0.05 & -0.296 & 1.72e-6 & 0.720 & 2.95e-5 & 0.629 & 8.53e-4 & 0.502 & 0.19 & 0.198 & 4.39e-4 & 0.528 \\
		poi-2.0$\Rightarrow$xerces-1.2 & 5.98e-8 & 0.816 & 0.002 & -0.458 & 7.37e-11 & 0.980 & 9.25e-9 & 0.864 & 3.01e-11 & 1 & 1.07e-7 & 0.800 & 2.74e-11 & 1 \\
		xalan-2.4$\Rightarrow$xerces-1.2 & 2.97e-11 & 1 & 3.25e-4 & 0.541 & 4.62e-8 & 0.822 & 7.57e-7 & 0.744 & 0.61 & -0.078 & 3.47e-9 & 0.889 & 2.43e-10 & 0.952 \\
		camel-1.0$\Rightarrow$velocity-1.6 & 0.044 & -0.304 & 4.25e-10 & 0.936 & 0.009 & 0.396 & 2.60e-10 & 0.951 & 0.115 & -0.238 & 1.99e-6 & 0.716 & 0.09 & 0.256 \\
		ivy-1.4$\Rightarrow$velocity-1.6 & 3.79e-7 & 0.764 & 5.45e-11 & 0.973 & 5.41e-10 & 0.933 & 4.05e-11 & 0.993 & 6.49e-4 & 0.513 & 1.25e-9 & 0.913 & 2.57e-11 & 0.996 \\
		tomcat$\Rightarrow$velocity-1.6 & 5.73e-5 & 0.606 & 0.589 & -0.082 & 8.96e-11 & 0.964 & 6.68e-11 & 0.982 & 5.09e-11 & 0.984 & 4.76e-7 & 0.758 & 2.58e-11 & 0.998 \\
		prop-6$\Rightarrow$velocity-1.6 & 9.73e-9 & 0.861 & 9.44e-10 & 0.920 & 3.77e-5 & 0.620 & 1.43e-8 & 0.853 & 3.28e-6 & 0.700 & 2.59e-7 & 0.771 & 0.877 & -0.024 \\
		poi-2.0$\Rightarrow$velocity-1.6 & 2.53e-4 & -0.551 & 8.04e-11 & 0.978 & 1.11e-7 & 0.798 & 1.95e-10 & 0.958 & 2.96e-11 & 1 & 0.739 & 0.051 & 1.11e-6 & 0.732 \\
		xalan-2.4$\Rightarrow$velocity-1.6 & 0.006 & 0.410 & 9.20e-9 & -0.864 & 2.08e-10 & 0.956 & 3.66e-11 & 0.996 & 1.66e-8 & 0.844 & 3.36e-9 & 0.889 & 2.77e-11 & 1 \\
		xerces-1.2$\Rightarrow$velocity-1.6 & 0.363 & -0.138 & 1.08e-8 & 0.856 & 0.021 & -0.347 & 1.89e-4 & 0.562 & 0.08 & 0.264 & 0.003 & 0.444 & 0.004 & -0.429 \\
		
		\hline
	\end{tabular}	
\end{table}

\section{Discussion}\label{sec:discussion}
In this section, we mainly study the following two problems: the effect of the parameter $\lambda$ and $\sigma$ on the performance of TOMOFWTNB.


\subsection{The effect of the parameter $\lambda$ on the performance of TOMOFWTNB?}

By default, we set $\lambda$ $\left(\lambda\in\left[0,1\right]\right)$ to 0.4. To study the effect of $\lambda$ on the performance of our proposed TOMOFWTNB, we perform experiments with different values of $\lambda$, i.e., 0, 0.1, ..., 1. Similar to the setting in RQ1-3, total of 34 groups of experiments are conducted for each $\lambda$ value. However, all instances of source dataset are used as training data here. Therefore, TOMOFWTNB is run one time on each source dataset, and then the average G-Measure and MCC values across all 34 experiments are reported.

\begin{figure*}[htbp]
	\centering
	\subfigure[The error-bars of G-Measure]{\label{disc1:GM}
		\includegraphics[width=0.49\textwidth]{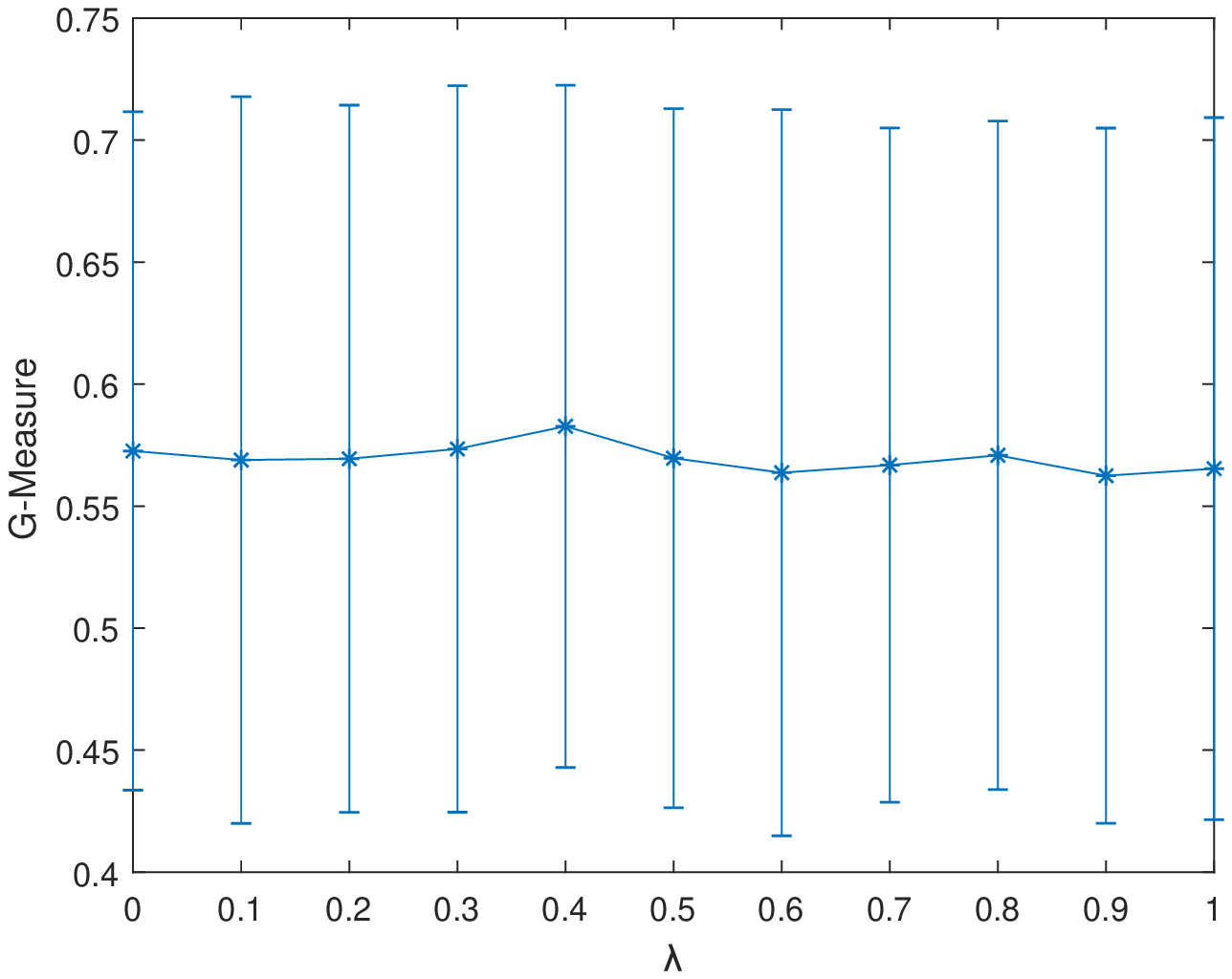}}	
	\subfigure[The error-bars of MCC]{\label{disc1:MCC}
		\includegraphics[width=0.49\textwidth]{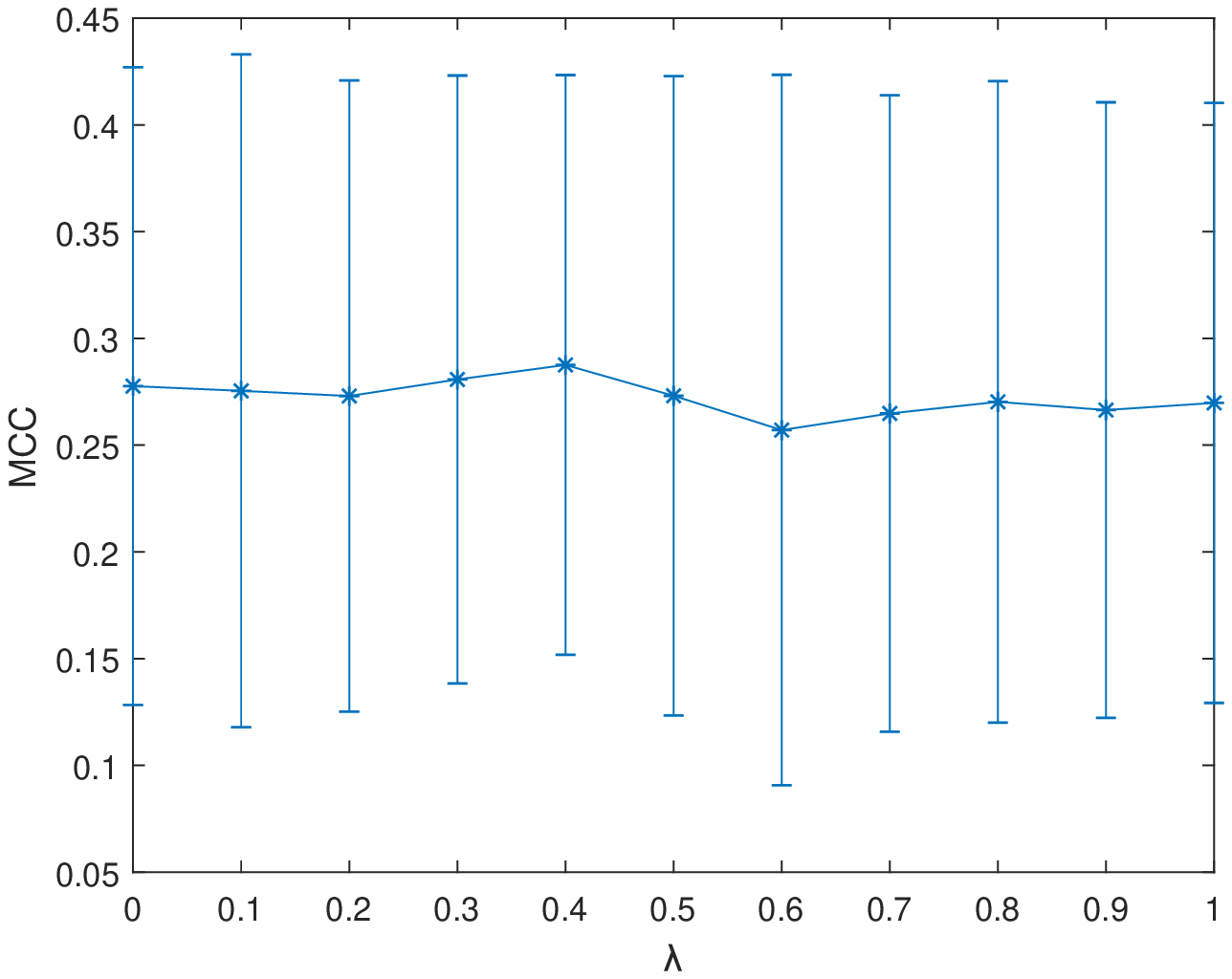}}
	\caption{The error-bars of G-Measure and MCC for TOMOFWTNB with different values of $\lambda$. The error-bar represents one standard deviation above and below the mean value.}
	\label {fig:disc-lamda0}
\end{figure*}

Figure \ref{fig:disc-lamda0} respectively shows the error-bars of G-Measure and MCC for TOMOFWTNB with different $\lambda$ values. The x-axis represents the value of $\lambda$, which denotes the weight of information of source data and $1-\lambda$ represents the weight of target data. The y-axis denotes the value of corresponding performance. The dot on the middle of the vertical line denotes the average performance of TOMOFWTNB on the 34 groups of experiments. From the figure, we notice that (1) if just the information of source data is taken into consideration (i.e., $\lambda=1$), TOMOFWTNB generally has the worst performance in terms of G-Measure and MCC; (2) from the view of overall trend, the performance of TOMOFWTNB generally become better when paying more and more attention to the information of unlabeled target data (i.e., along with the increasing of $1-\lambda$); (3) both G-Measure and MCC of TOMOFWTNB are not very sensitive to the value of $\lambda$; (4) when $\lambda$ is in the interval of [0,0.5], TOMOFWTNB has better performance. Therefore, in the experiments, $\lambda$=0.4, which can be utilized as the default setting in practice.



\subsection{The effect of the parameter $\sigma$ on the performance of TOMOFWTNB?}
Another parameter of TOMOFWTNB is $\sigma$ $\left(\sigma>0\right)$. By default, we set $\sigma$ to 1. To study the effect of $\sigma$ on the performance of our proposed TOMOFWTNB, we perform experiments with different $\sigma$ values, i.e., 0.2, 0.4, ..., 2. Here, the same evaluation settings as above section is applied.

\begin{figure*}[htbp]
	\centering
	\subfigure[The error-bars of G-Measure]{\label{disc2:GM}
		\includegraphics[width=0.49\textwidth]{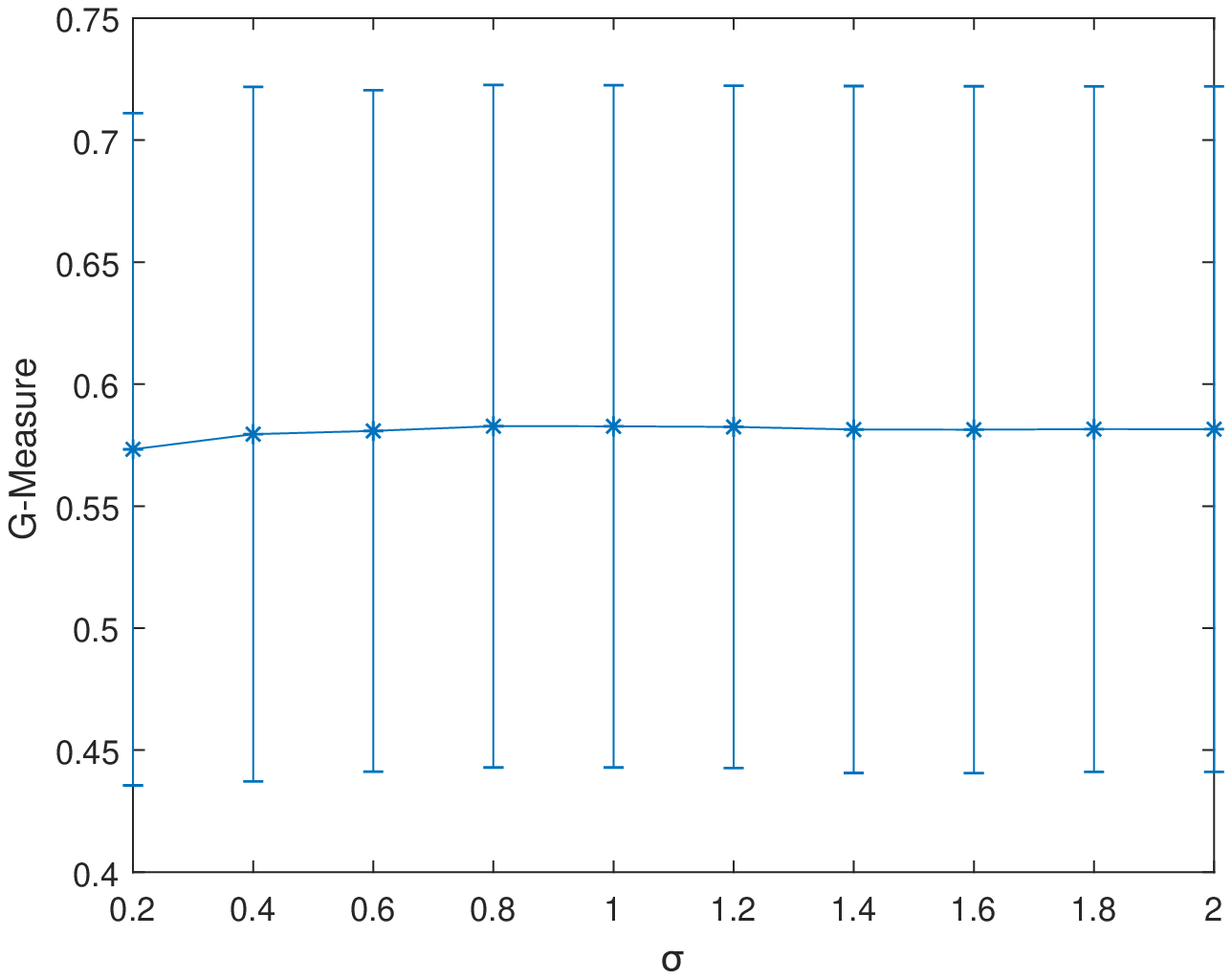}}	
	\subfigure[The error-bars of MCC]{\label{disc2:MCC}
		\includegraphics[width=0.49\textwidth]{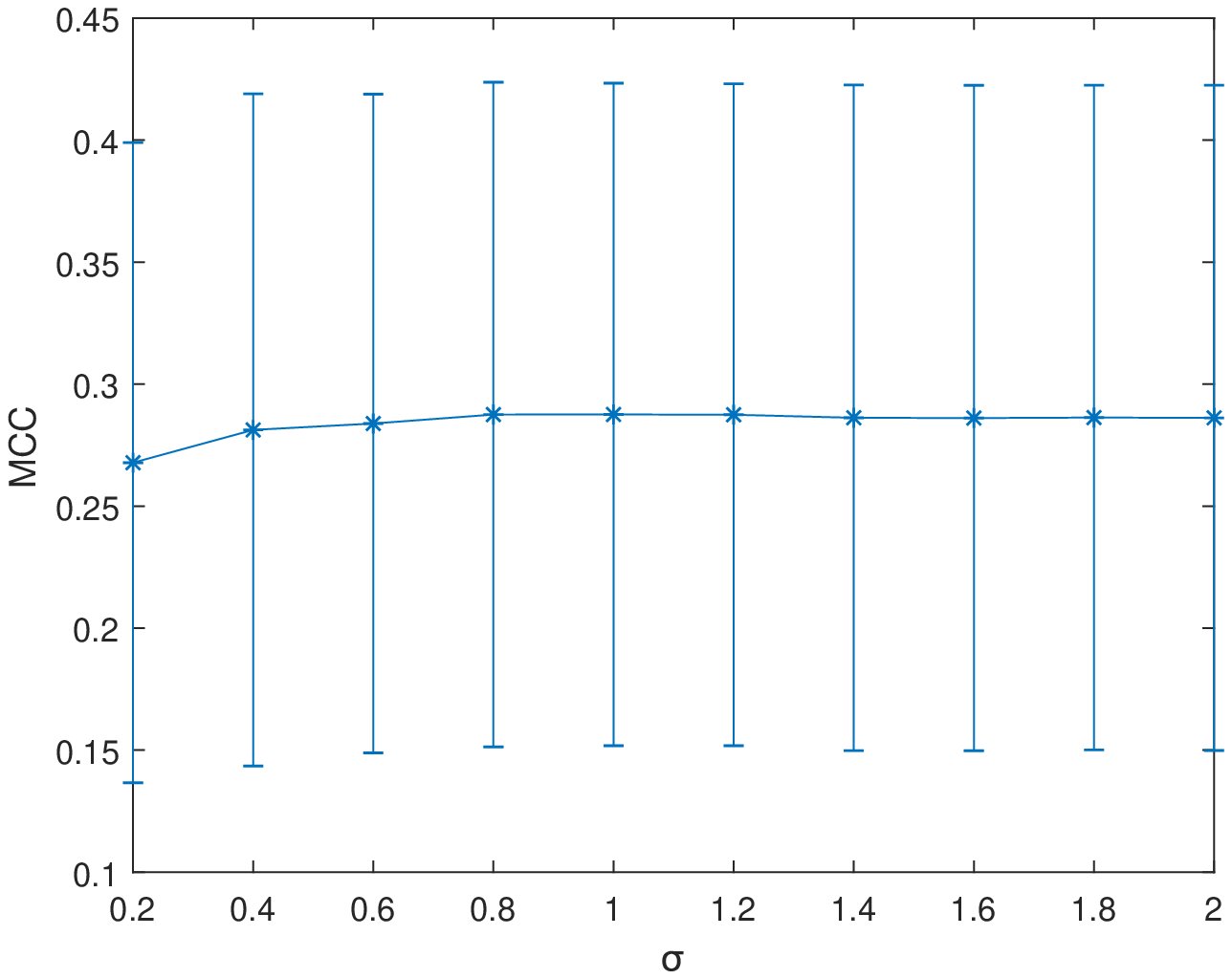}}
	\caption{The error-bars of G-Measure and MCC for TOMOFWTNB obtained from 34 groups of experiments with different values of $\sigma$}
	\label {fig:disc-sigma}
\end{figure*}

Figure \ref{fig:disc-sigma} shows the error-bars of G-Measure and MCC for TOMOFWTNB with different $\sigma$ values. The x-axis represents the value of $\sigma$. The y-axis denotes the value of corresponding performance. The dot on the middle of the vertical line denotes the average performance of TOMOFWTNB across the 34 groups of experiments. From the figure, we notice that (1) when $\sigma$ ranging from 0.2 to 0.8, the performance of TOMOFWTNB is becoming slightly better in terms of both G-Measure and MCC; (2) since $\sigma=0.8$, the performance TOMOFWTNB is very stable in terms of G-Measure and MCC. It indicates that TOMOFWTNB is not sensitive to the value of $\sigma$. Therefore, with respect to $\sigma=1$, it can be used as the default setting in practice.


\section{Threats to Validity}\label{sec:threats}
In this section, some potential threats to the validity of our research are presented.

\subsection{Internal Validity}

Threats to the internal validity of our study mainly relate to re-implementation of baselines and the analysis of causal effect between independent and dependent variables. Since the authors did not provide the source code of baselines except CTKCCA\footnote{ https://sites.google.com/site/cstkcca/}, we implement other baselines carefully according to the description in corresponding studies. Although we have checked our source code carefully, there still may be errors that we did not notice. Since the analysis of the causal effect is not the objective of this study and the benchmark datasets used in our study have been widely used in previous SDP studies \cite{Xia2016HYDRA,Chen2015Negative,He2015An,He2013Learning}, we did not analyze this effect. Related works can be seen in \cite{Gyimothy2005Empirical,Olague2007Empirical}. Therefore, this threat also exists in our study.

\subsection{External Validity}

External validity means the degree to generalize the research results to other situations \cite{Xia2016HYDRA,Malhotra2017An}. In this study, to evaluate the performance of our proposed TOMO, FWTNB, and TOMOFWTNB, experiments are performed on 11 public defect datasets collected from different software projects. The prediction performance is evaluated in terms of two performance evaluation measures (i.e., G-Measure and MCC). To demonstrate the superiority of TOMOFWTNB compared with existing CPDP methods, both Wilcoxon rank-sum test and Cliff's delta are used. However, we still cannot claim that the findings will be completely suitable for other contexts, e.g., different defect datasets. We should perform the proposed approach on more defect datasets to reduce this threat.



\subsection{Construct Validity}

Threats to construct validity for our study mainly refer to the bias of baselines and performance evaluation measures. Total of seven CPDP models were selected as baselines. These models include both instance-based CPDP models (such as Burak filter and TNB) and feature-based CPDP models (aka. heterogeneous defect prediction models, such as CCA+ and CTKCCA). Due to the space limitation, we cannot compare the proposed approach with all existing CPDP models in our study. In this study, two measures (G-Measure and MCC) were utilized to evaluate defect prediction performance. These measures have been widely used to evaluate the performance of defect prediction models when the datasets are imbalanced in previous SDP studies \cite{Peters2013Better,Chen2015Negative,He2013Learning,Peters2013Balancing,Zhang2016Towards,Bowes2016Mutation}. However, some other measures, such as precision, AUC, and F1, also have been used in some previous studies.

\section{Conclusion}\label{sec:conclusion}

Cross-project defect prediction (CPDP), which aims to build a prediction model historical defect data (source data) from other projects and apply the model on a new project data (target data), has drawn increasing attention of academic and industrial communities. Some CPDP approaches have been proposed and demonstrated the effectiveness. Distribution complexity and distribution differences between the source and target datasets limit the performance of CPDP methods. Software defect datasets have class-imbalance nature. However, there are few previous CPDP methods to address this problem. Furthermore, previous CPDP approaches usually assume that the importance of different software metrics is the same and just treat them equally. 

To address this, in this paper, we propose a transfer learning oriented minority over-sampling technique (TOMO) based feature weighting transfer naive Bayes (FWTNB) approach, i.e., TOMOFWTNB, for CPDP. TOMO is used to address the class-imbalance problem in CPDP scenario. FWTNB uses feature weighting strategy to obtain better CPDP performance.


We performed 34 group of experiments on 11 public defect datasets collected from different software projects. The prediction performance is evaluated in terms of G-Measure and MCC, which are effective on imbalanced data. Wilcoxon rank-sum test and Cliff's delta are used to demonstrate the superiority of TOMOFWTNB compared with existing CPDP methods. The experimental results show that The experimental results indicate that (1) TOMO is very effective for addressing the class-imbalance problem in CPDP scenario; (2) our feature weighting strategy is helpful for CPDP; (3) TOMOFWTNB significantly outperforms previous state-of-the-art both instance-based and feature-based CPDP approaches in most cases.


In the future, we plan to perform our model on more defect datasets. Furthermore, we will extend our transfer learning oriented synthetic over-sampling technique (TOMO) to heterogeneous defect prediction scenario.

To make it convenient for other researchers to replicate or verify our research, we open the benchmark datasets, the source code, and the original experimental results on the following website:
\url{}.



\section*{Acknowledgments}
 We would like to thank the editors and anonymous reviewers for their constructive comments and suggestions for the improvement of this paper.

\section*{References}

\bibliography{mybibfile}

\end{document}